\documentclass[prd,amsmath,amssymb,showpacs,superscriptaddress,nofootinbib,twocolumn]{revtex4}
\usepackage{multirow}
\usepackage{graphicx,epsfig}
\usepackage{dcolumn}
\usepackage{bm}
\usepackage{rotating}
\usepackage{color}
\usepackage{subfigure}
\usepackage{pstricks}
\usepackage{soul}
\usepackage{overpic}
\usepackage[english]{babel}
\def \jpsi      {J/\psi}

\def \etap      {\eta^{\prime}}
\def \chicj     {\chi_{cJ}}
\def \chiczero  {\chi_{c0}}
\def \chicone   {\chi_{c1}}
\def \chictwo   {\chi_{c2}}
\def \etapkk    {\eta'K^+K^-}
\def \etaptogammarho {\etap\to\gamma\rho^{0}}
\def \etaptoetapipi  {\etap\to\eta\pi^+\pi^-}
\def \ee         {e^+e^-}
\def \kk         {K^+K^-}
\def \GG         {\gamma\gamma}
\def \pp         {\pi^+\pi^-}
\def \kstar      {K^*_0(1430)}
\def \mev        {\,\mathrm{MeV}}
\def \mevcc      {\,\mathrm{MeV}/c^2}

\def \gevcc      {\,\mathrm{GeV}/c^2}
\begin{document}
\normalsize
\parskip=5pt plus 1pt minus 1pt

\title{\boldmath  Measurement of $\chi_{cJ}$ decaying into $\eta^{\prime}K^+K^-$}
\author{
M.~Ablikim$^{1}$, M.~N.~Achasov$^{8,a}$, X.~C.~Ai$^{1}$, O.~Albayrak$^{4}$, M.~Albrecht$^{3}$, D.~J.~Ambrose$^{41}$, F.~F.~An$^{1}$, Q.~An$^{42}$, J.~Z.~Bai$^{1}$, R.~Baldini Ferroli$^{19A}$
% ...
\\
\vspace{0.2cm}
(BESIII Collaboration)\\
\vspace{0.2cm} {\it
$^{1}$ Institute of High Energy Physics, Beijing 100049, People's Republic of China\\
% ...
$^{3}$ Bochum Ruhr-University, D-44780 Bochum, Germany\\
$^{4}$ Carnegie Mellon University, Pittsburgh, Pennsylvania 15213, USA\\
% ...
$^{8}$ G.I. Budker Institute of Nuclear Physics SB RAS (BINP), Novosibirsk 630090, Russia\\
% ...
$^{19}$ (A)INFN Laboratori Nazionali di Frascati, I-00044, Frascati, Italy; (B)INFN and University of Perugia, I-06100, Perugia, Italy\\
% ...
$^{41}$ University of Rochester, Rochester, New York 14627, USA\\
$^{42}$ University of Science and Technology of China, Hefei 230026, People's Republic of China\\
% ...
$^{a}$ Also at the Novosibirsk State University, Novosibirsk, 630090, Russia\\
% ...
}
}

\date{\today}

\begin{abstract}
Using $(106.41\pm 0.86) \times 10^{6}$ $\psi(3686)$ events collected
with the BESIII detector at BEPCII, we study
for the first time the decay $\chi_{cJ}\to\eta'K^+K^-$ ($J=1, 2$), where $\eta'\to\gamma\rho^{0}$ and
$\eta'\to\eta\pi^+\pi^-$.  A partial wave analysis in the covariant tensor
amplitude formalism is performed for the decay
$\chi_{c1}\to\eta'K^+K^-$. Intermediate processes $\chi_{c1}\to\eta'
f_0(980)$, $\chi_{c1}\to\eta' f_0(1710)$, $\chi_{c1}\to\eta' f_2'(1525)$
and $\chi_{c1}\to K^*_0(1430)^{\pm}K^{\mp}$ ($K^*_0(1430)^{\pm}\to\eta'
K^{\pm}$) are observed with statistical significances larger than
5$\sigma$, and their branching fractions are measured.
\end{abstract}

\maketitle
\section{\boldmath{Introduction}}
Exclusive heavy quarkonium decays provide an important laboratory for
investigating perturbative Quantum Chromodynamics (pQCD). Compared to
$\jpsi$ and $\psi(3686)$ decays, relatively little is known concerning
$\chicj$ decays~\cite{pdg}. More experimental data on exclusive decays
of $P$-wave charmonia are important for a better understanding of the
decay dynamics of the $\chicj$ ($J$=0, 1, 2) states, as well as
testing QCD based calculations.
Although these $\chicj$ states are not directly produced in $\ee$ collisions,
they are produced copiously in $\psi(3686)$ $E1$ transitions, with
branching fractions around 9\%~\cite{pdg} each.
The large $\psi(3686)$ data sample taken with the Beijing Spectrometer (BESIII)
located at the Beijing Electron-Positron Collider (BEPCII) provides an
opportunity for a detailed study of $\chicj$ decays.

QCD theory allows the existence of glueballs, and glueballs are
expected to mix strongly with nearby conventional $q\bar{q}$
states~\cite{VVAnisovich}.  For hadronic decays of the $\chicone$, two-gluon
annihilation in pQCD is suppressed by the Landau-Yang
theorem~\cite{LandauYang} in the on-shell limit. As a result, the
annihilation is expected to be dominated by the pQCD hair-pin diagram.
The decay $\chi_{c1} \to PS$, where $P$ and $S$ denote a pseudoscalar
and a scalar meson, respectively, is expected to be sensitive to the
quark contents of the final-state scalar meson. And by tagging the quark contents
 of the recoiling pseudo-scalar meson, the process can be used in testing the
glueball-$q\overline{q}$ mixing relations among the scalar mesons $S$,
{\it i.e.} $f_0(1370)$, $f_0(1500)$, $f_0(1710)$.
A detailed calculation can be found in Ref.~\cite{wangqian}.

The $K^*_0(1430)$ state is perhaps the least controversial of the
light scalar isobar mesons~\cite{pdg}. Its properties are still
interesting since it is highly related to the lineshape of the
controversial $\kappa$ meson ($K\pi$ $S$-wave scattering at mass
threshold) in various studies. Until now, $K^*_0(1430)$ has been
observed in $\kstar\to K\pi$ only, but it is also expected to couple to
$\eta'K$ ~\cite{cleokstar,buggkstar}. The opening of
the $\eta'K$ channel will affect its lineshape. $\chicone\to\etapkk$
is a promising channel to search for $K^*_0(1430)$ and study its
properties. The decays $\chi_{c0,2}\to\kstar K$ are
forbidden by spin-parity conservation.

In this paper, we study the decay $\chicj\to\etapkk$ with $\etaptogammarho$
(mode I) and $\etaptoetapipi, \eta\to\gamma\gamma$ (mode II). Only
results for $\chicone$ and $\chictwo$ are given, because
$\chiczero\to\etapkk$ is forbidden by spin-parity conservation.  A
partial wave analysis (PWA) in the covariant tensor amplitude
formalism is performed for the process $\chicone$, and results on
intermediate processes involved are given.  For $\chictwo\to\etapkk$, due to low statistics,
 a simple PWA is performed, and the result is used to estimate the event selection efficiency.
 The data sample used in this analysis consists of 156.4 pb$^{-1}$ of
 data taken at $\sqrt{s} = 3.686 \;\mathrm{GeV}/c^{2}$ corresponding to (106.41$\pm$0.86$)\times 10^6$
$\psi(3686)$ events~\cite{totaln}.

\section{\boldmath{Detector and Monte-Carlo simulation}}
BESIII~\cite{liu2} is a general purpose detector at the
BEPCII accelerator for studies of hadron spectroscopy as well as
$\tau$-charm physics~\cite{liu4}. The design peak luminosity of the
double-ring $e^{+}e^{-}$ collider, BEPCII, is 10$^{33}$
cm$^{-2}$s$^{-1}$ at center-of-mass energy of 3.78 GeV. The BESIII
detector with a geometrical acceptance of 93\% of 4$\pi$, consists of
the following main components: 1) a small-cell, helium-based main
drift chamber (MDC) with 43 layers, which measures
tracks of charged particles and provides a measurement of the
specific energy loss $dE/dx$.
The average single wire resolution
is 135 $\mu$m, and the momentum resolution for 1 GeV/$c$ charged
particles in a 1 T magnetic field is 0.5\%; 2) an electromagnetic
calorimeter (EMC) consisting of 6240 CsI(Tl) crystals arranged in a
cylindrical shape (barrel) plus two end-caps. For 1.0 GeV/$c$ photons,
the energy resolution is 2.5\% (5\%) in the barrel (endcaps), and the
position resolution is 6 mm (9 mm) in the barrel (end-caps); 3) a
Time-Of-Flight system (TOF) for particle identification (PID) composed
of a barrel part constructed of two layers with 88 pieces of 5 cm
thick, 2.4 m long plastic scintillators in each layer, and two endcaps
with 48 fan-shaped, 5 cm thick, plastic scintillators in each
endcap. The time resolution is 80 ps (110 ps) in the barrel (endcaps),
corresponding to a $K/\pi$ separation by more than $2\sigma$ for
momenta below about 1 GeV/$c$; 4) a muon chamber system (MUC) consists
of 1000 m$^{2}$ of Resistive Plate Chambers (RPC) arranged in 9 layers
in the barrel and 8 layers in the end-caps and incorporated in the
return iron yoke of the superconducting magnet. The position
resolution is about 2 cm.

The optimization of the event selection and the estimation of
backgrounds are performed through Monte Carlo (MC) simulation. The
\textsc{geant}{\footnotesize 4}-based simulation software
\textsc{boost}~\cite{liu5} includes the geometric and material
description of the BESIII detectors and the detector response and
digitization models, as well as the tracking of the detector running
conditions and performance. The production of the $\psi(3686)$ resonance is
simulated by the MC event generator \textsc{kkmc}~\cite{liu6}, while
the decays are generated by \textsc{evtgen}~\cite{liu7} for known
decay modes with branching fractions being set to world average
values~\cite{pdg}, and by \textsc{lundcharm}~\cite{liu9} for the
remaining unknown decays.

\section{\boldmath{Event selection}}
The final states of the sequential decay $\psi(3686)\to\gamma\chicj$,
$\chicj\to\etap\kk$ have the topologies $\gamma\gamma\kk\pp$ or
$\gamma\gamma\gamma\kk\pp$ for $\etap$ decay modes I or II,
respectively. Event candidates are required to have four charged
tracks and at least two (three) good photons for mode I (II).

Charged tracks in the polar angle range $|\cos\theta|<0.93$ are
reconstructed from MDC hits. The closest point to the beamline of each
selected track should be within $\pm$10 cm of the interaction point in
the beam direction, and within 1 cm in the plane perpendicular to the beam.
The candidate events are required to have four well reconstructed
charged tracks with net charge zero. TOF and $dE/dx$ information is combined
to form particle identification (PID) confidence levels for the $\pi$, $K$
and $p$ hypotheses. Kaons are identified by requiring the PID probability
($Prob$) to be $Prob(K)>Prob(\pi)$ and $Prob(K)>Prob(p)$. Two identified
kaons with opposite charge are required. The other two charged tracks
are assumed to be pions.

Photon candidates are reconstructed by clustering signals in EMC crystals.
The photon candidates in the barrel ($|\cos\theta|<0.80$) of the EMC
are required to have at least $25\mev$ total energy deposition, or in
the endcap ($0.86<|\cos\theta|<0.92$) at least $50\mev$ total energy
deposition, where $\theta$ is the polar angle of the shower. The photon
candidates are further required to be isolated from all charged tracks
by an angle $> 5^\circ$ to suppress showers from charged particles.
Timing information from the EMC is used to suppress electronic noise
and energy deposition unrelated to the event.

A four-constraint (4C) energy-momentum conserving kinematic fit is
 applied to candidate events under the $\gamma\gamma(\gamma)\kk\pp$
 hypothesis.  For events with more than two (three) photon candidates,
 all of the possible two (three) photon combinations are fitted, and
 the candidate combination with the minimum $\chi^2_{4C}$ is selected, and
 it is required that $\chi^2_{4C}<40$ $(50)$.

\begin{figure*}[htbp]
\centering
\begin{overpic}[width=8.0cm,height=5.0cm,angle=0]{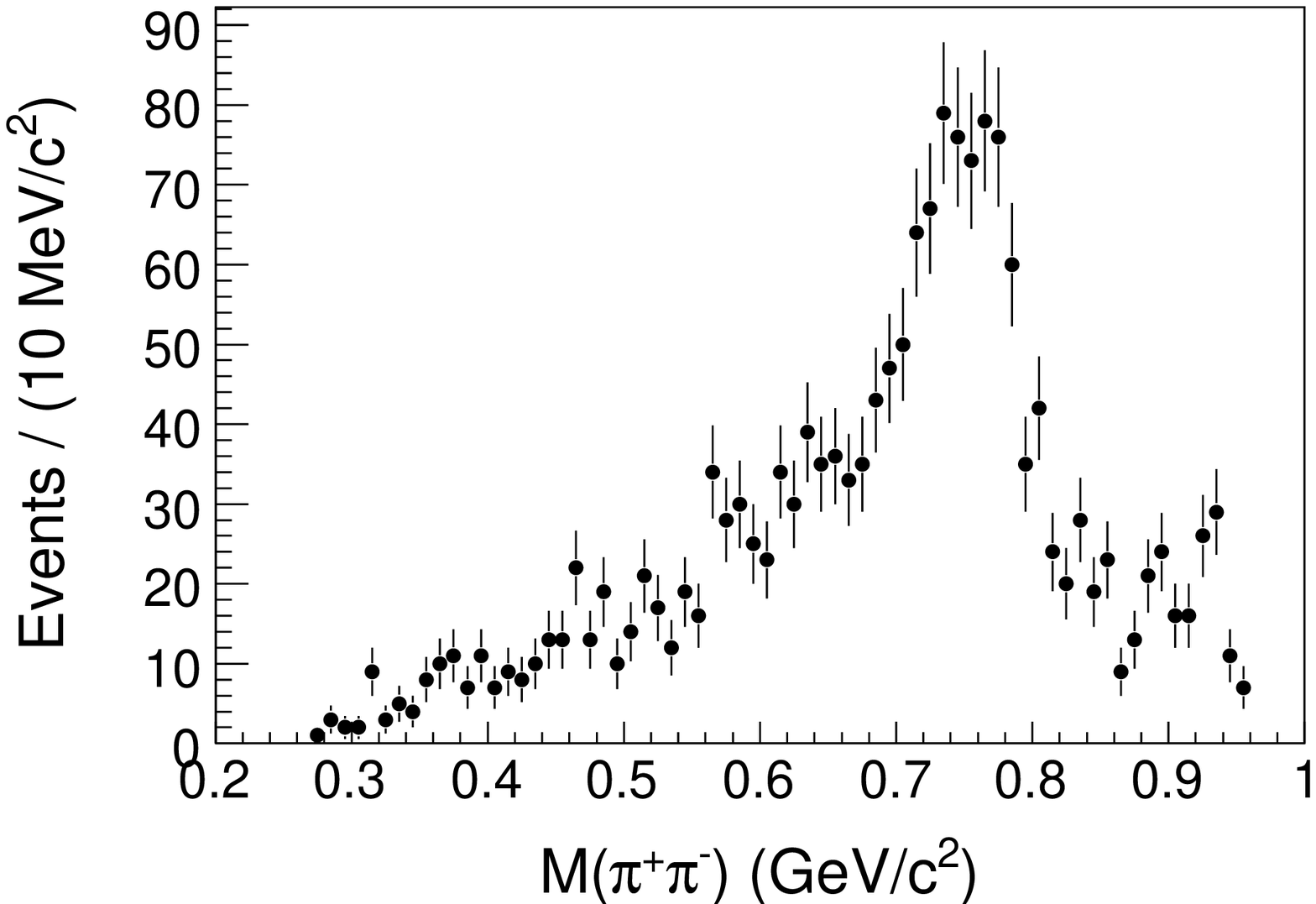}
\put(85,50){\large\bf (a)}
\end{overpic}
\begin{overpic}[width=8.0cm,height=5.0cm,angle=0]{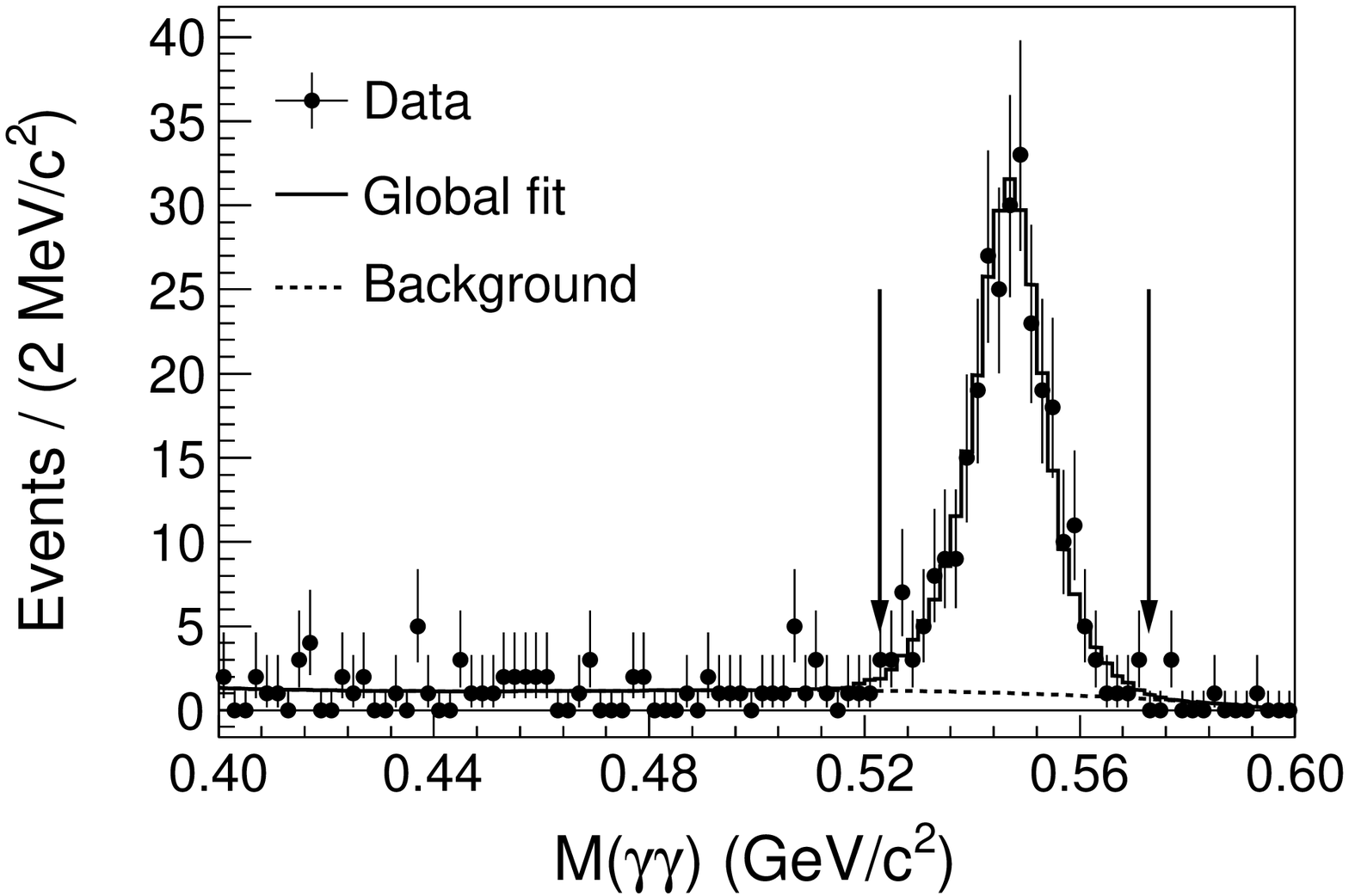}
\put(85,50){\large\bf (b)}
\end{overpic}
\vskip -0.3cm
\parbox[1cm]{16cm} {
\caption{The invariant mass distributions of (a) $\pp$ in mode I, and
(b) $\gamma\gamma$ in mode II. The arrows show the $\eta$ signal
region.}
 \label{FigGG}
}
\end{figure*}

In the $\etap$ decay mode I, the photon with the smaller
$|M(\gamma\pp)-M(\etap)|$ is assigned as the photon from $\etap$
decay, and the other one is tagged as the photon from the radiative
decay of $\psi(3686)$. The mass requirement
$|M(\gamma\gamma)-M(\pi^0)|>15\mevcc$ is applied to remove backgrounds
with $\pi^0$ in the final
state.  $|M(\pp)_{rec}-M(\jpsi)|>8\mevcc$ and
$|M(\gamma\gamma)_{rec}-M(\jpsi)|>22\mevcc$ are further used to
suppress backgrounds from $\psi(3686)\to\pp\jpsi$ with
$\jpsi\to(\gamma/\pi^0/\gamma\pi^0)\kk$, as well as from
$\psi(3686)\to\gamma\chicj\to\gamma \gamma\jpsi$ or
$\psi(3686)\to(\eta/\pi^0)\jpsi$ with $\jpsi\to\kk\pp$, where
$M(\pp)_{rec}$ and $M(\gamma\gamma)_{rec}$ are the recoil masses from
the $\pp$ and $\gamma\gamma$ systems,
respectively. Figure~\ref{FigGG}(a) shows the invariant mass
distribution of $\pp$, and a clear $\rho^{0}$ signal is observed.
For the $\etap$ decay mode II, candidate events are rejected if any pair
of photons has $|M(\GG)-M(\pi^0)|<20\mevcc$, in order to suppress
backgrounds with $\pi^0$ in the final state.  The $\eta$ candidate
is selected as the photon pair whose invariant mass is closest to
the $\eta$ mass~\cite{pdg}.  The $M(\GG)$ distribution, shown in
Fig.~\ref{FigGG}(b), is fitted with the MC simulated $\eta$ signal
shape plus a $3^{rd}$ order polynomial background function.
$|M(\GG)-M(\eta)|<25\mevcc$ is required to select the $\eta$
signal.

\begin{figure*}[htbp]
\centering
\begin{overpic}[width=8.0cm,height=5.0cm,angle=0]{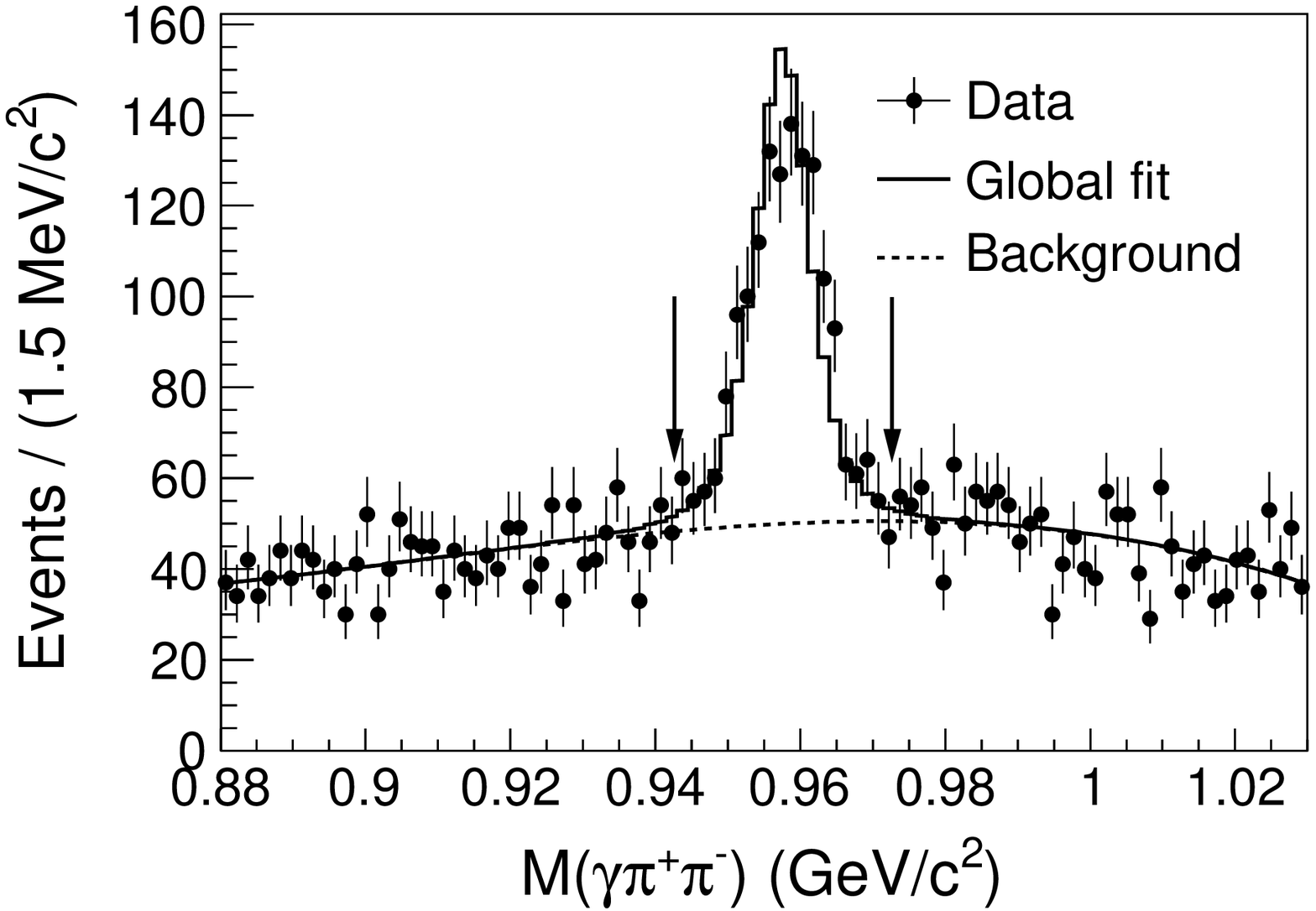}
\put(85,50){\large\bf (a)}
\end{overpic}
\begin{overpic}[width=8.0cm,height=5.0cm,angle=0]{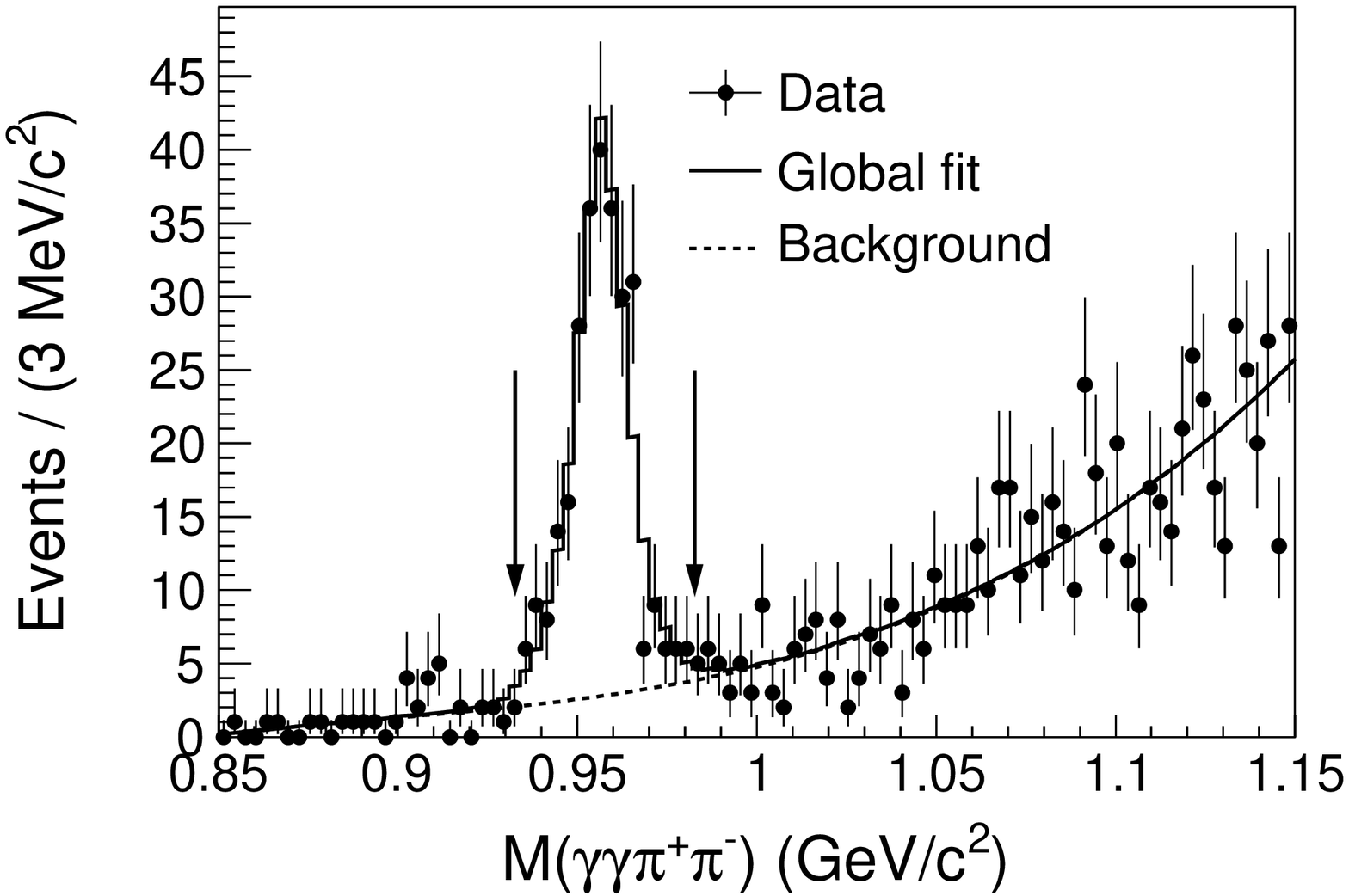}
\put(85,50){\large\bf (b)}
\end{overpic}
\vskip -0.3cm
\parbox[1cm]{16cm} {
\caption{The invariant mass distributions of (a) $\gamma\pp$ in the
decay mode I, and (b) $\GG\pp$ in the decay mode II. The arrows show the
$\etap$ signal region.}
 \label{etaandetaprime}
}
\end{figure*}

After the above event selection, the invariant mass distributions of
$\gamma\pp$ and of $\gamma\gamma\pp$ in the two $\etap$ decay modes
are shown in Fig.~\ref{etaandetaprime}.  The $\etap$ signals are seen
clearly, and the distributions are fitted with the MC simulated
$\etap$ signal shape plus a $3^{rd}$ order polynomial function for the
background.  $|M(\gamma\pp)-M(\etap)|<15\mevcc$ and
$|M(\eta\pp)-M(\etap)|<25\mevcc$ are used to select the $\etap$
signal in the two decay modes, respectively.

\section{\boldmath{Background studies}}
\begin{figure*}[htbp]
\centering
\begin{overpic}[width=8.0cm,height=5.0cm,angle=0]{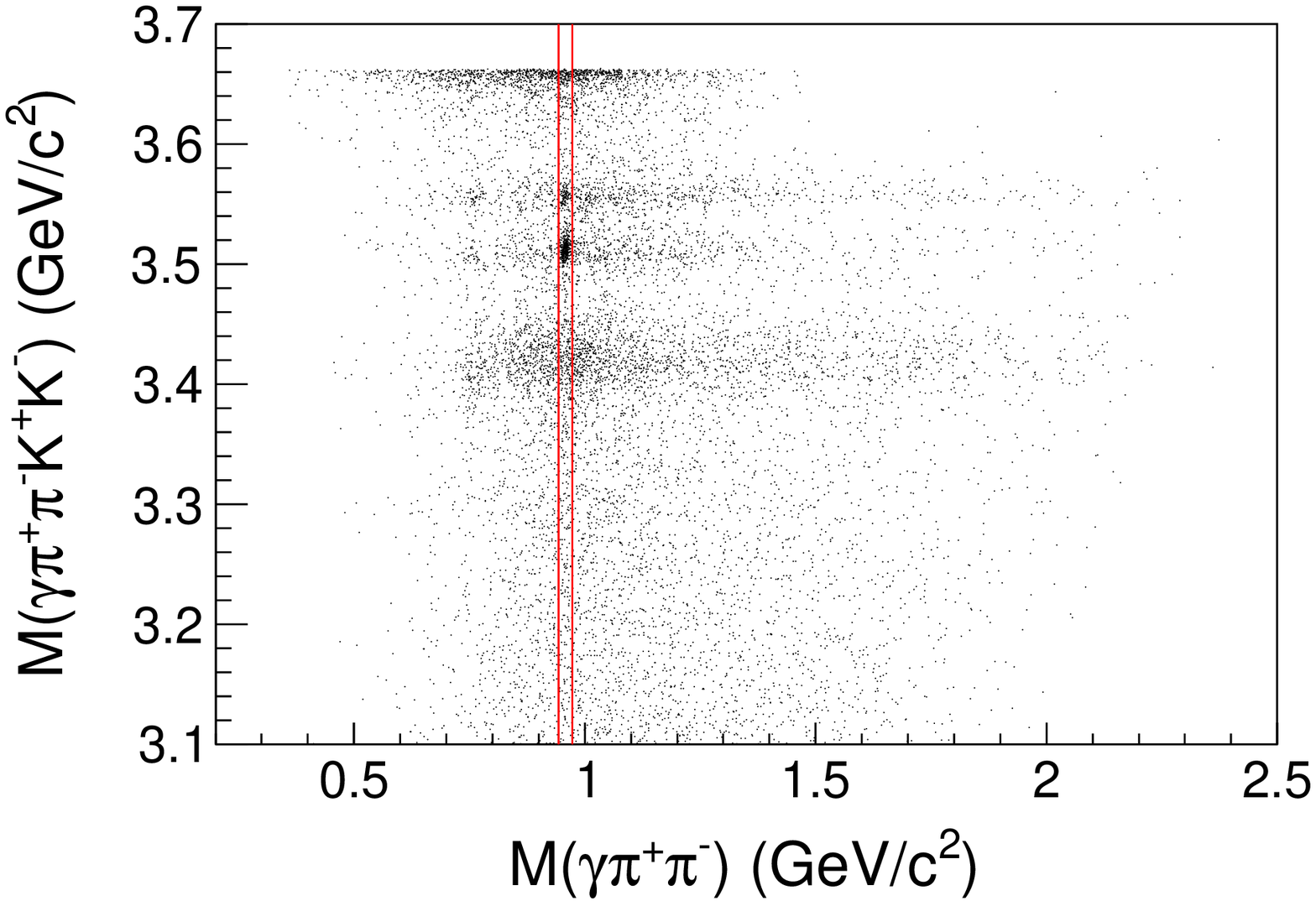}
\put(80,50){\large\bf (a)}
\end{overpic}
\begin{overpic}[width=8.0cm,height=5.0cm,angle=0]{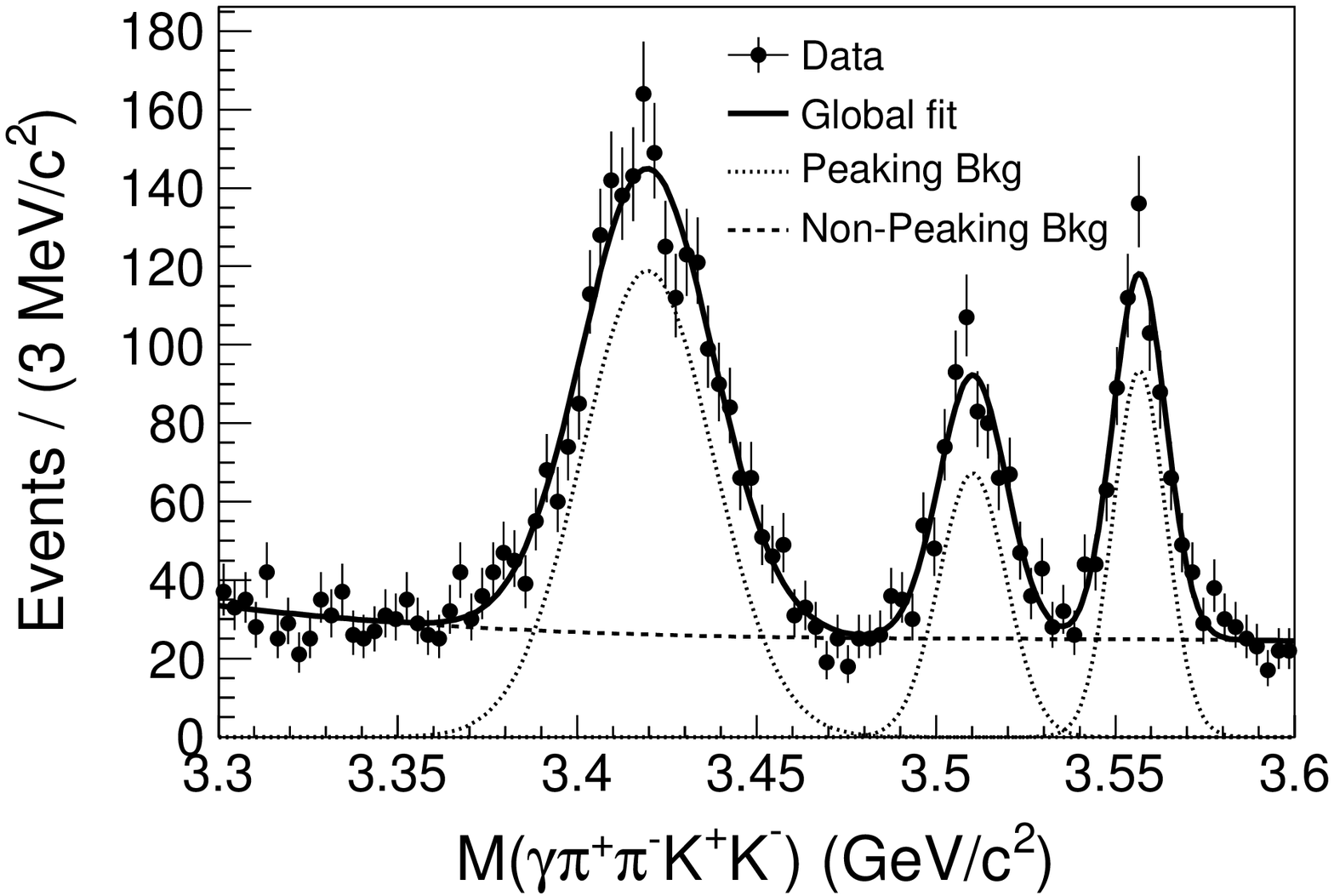}
\put(80,50){\large\bf (b)}
\end{overpic}
\parbox[1cm]{16cm} {
\caption{(color online) (a) The scatter plot of $M(\gamma\pp\kk)$ versus $M(\gamma\pp)$.
          The two vertical lines show the $\etap$ signal region.
         (b) The $\gamma\pp\kk$ invariant mass of events with
         $M(\gamma\pp)$ outside the $\etap$ range
         in the $\etap$ decay mode I.}
 \label{2Detapsideband}
}
\end{figure*}

\begin{figure*}[htbp]
\centering
\begin{overpic}[width=8.0cm,height=5.0cm,angle=0]{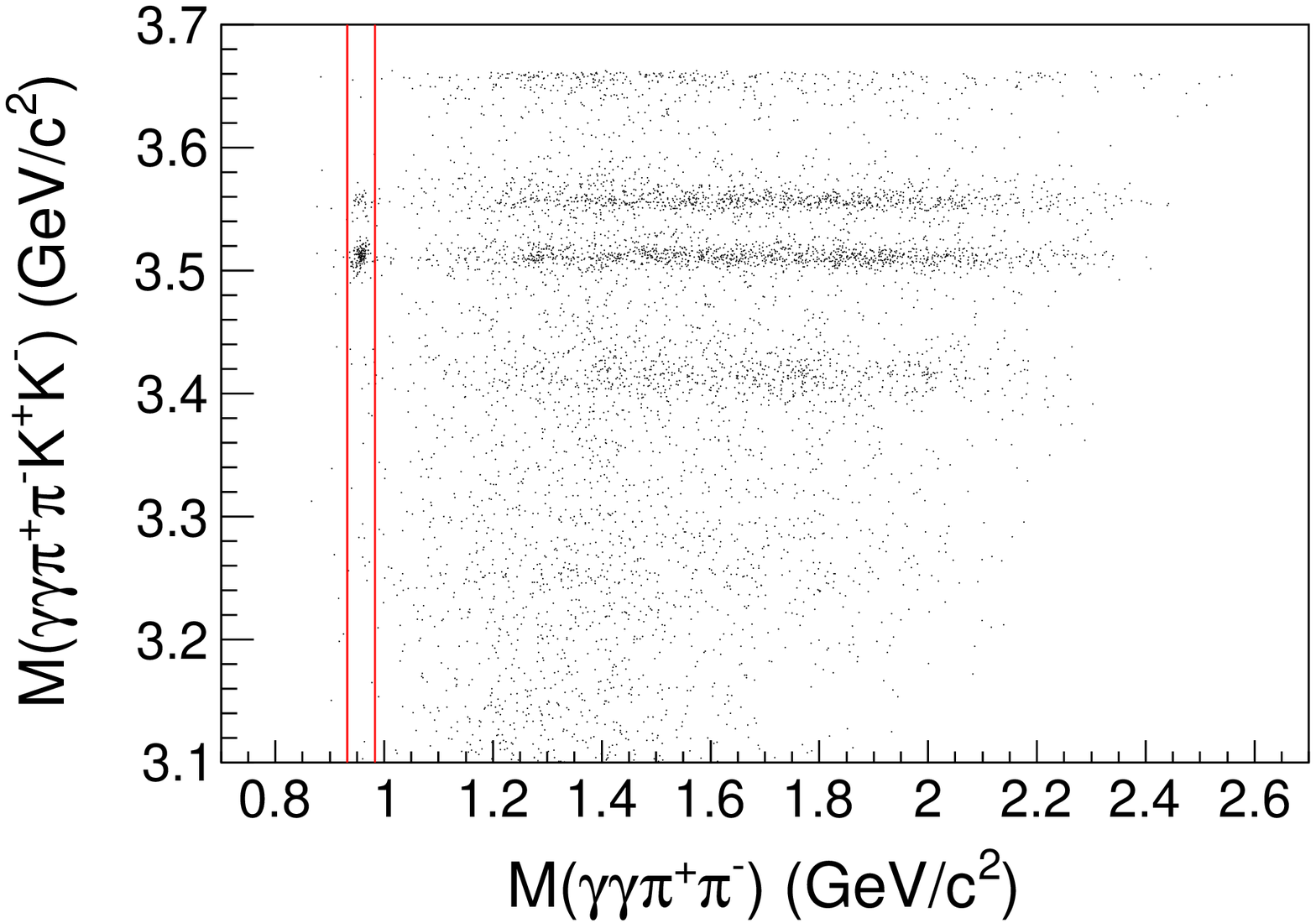}
\put(80,50){\large\bf (a)}
\end{overpic}
\begin{overpic}[width=8.0cm,height=5.0cm,angle=0]{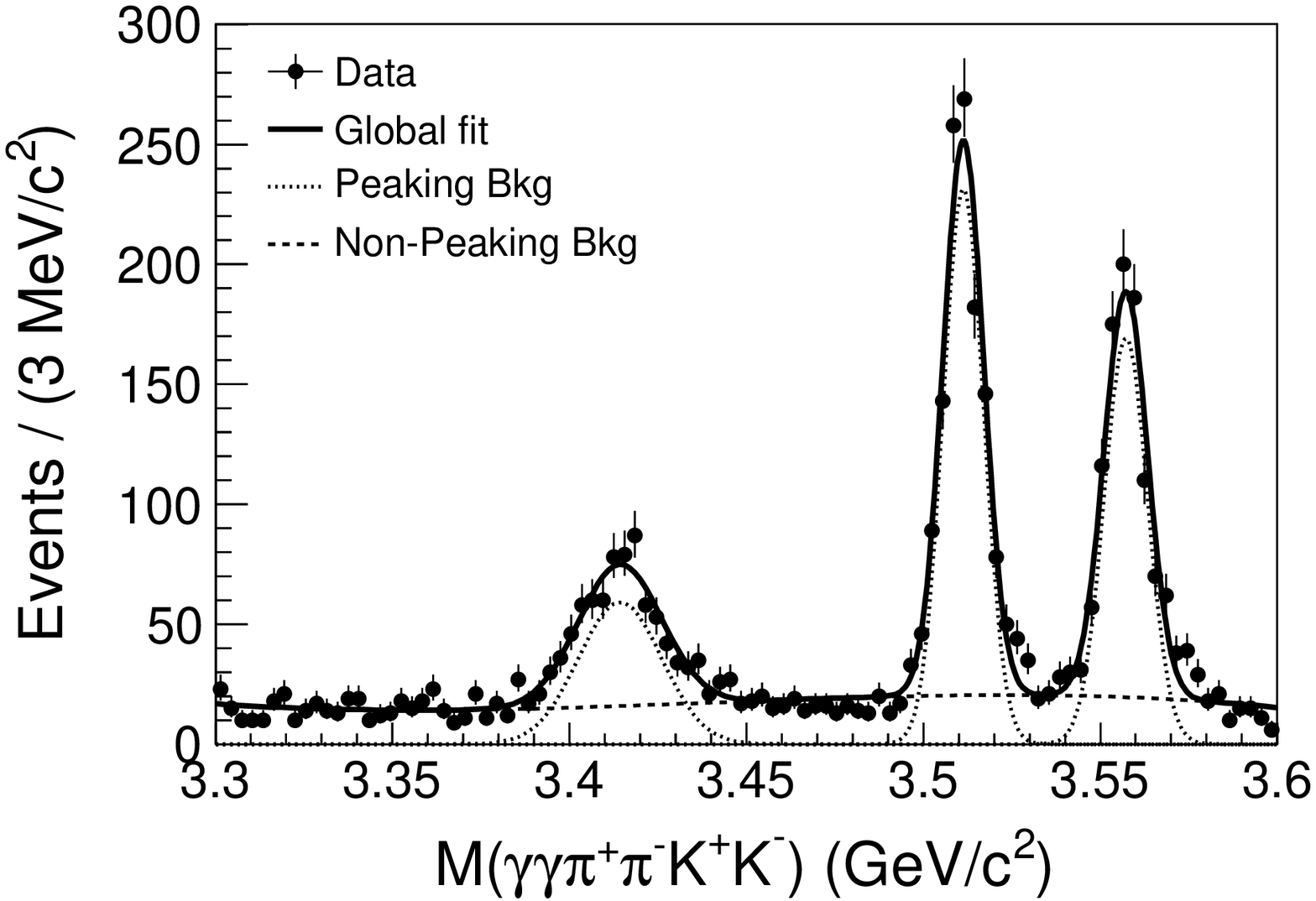}
\put(80,50){\large\bf (b)}
\end{overpic}
\parbox[1cm]{16cm} {
\caption{(color online) (a) The scatter plot of $M(\gamma\gamma\pp\kk)$ versus
         $M(\gamma\gamma\pp)$ distribution. The two vertical lines show the $\etap$ signal region. (b) The
         $\gamma\gamma\pp\kk$ invariant mass of events with
         $M(\gamma\gamma\pp)$ outside the $\etap$ range in the $\etap$
         decay mode II.}
\label{sidebandshape}
}
\end{figure*}

The scatter plots of the invariant mass of $\gamma(\gamma)\pp\kk$
versus that of $\gamma(\gamma)\pp$ are shown in
Fig.~\ref{2Detapsideband}(a) (mode I) and Fig.~\ref{sidebandshape}(a)
(mode II), respectively.  Two clusters of events in the $\chi_{c1,2}$
and $\etap$ signal regions, which arise from the signal processes of
$\psi(3686)\to\gamma\chi_{c1,2}$, $\chi_{c1,2}\to\etapkk$, are
clearly visible.
Clear $\chicj$ bands are also observed outside the $\etap$ signal
region.

Inclusive and exclusive MC studies are carried out to investigate
potential backgrounds.  The dominant backgrounds are found to be
$\psi(3686)\to\gamma\chicj$,
$\chicj\to\kk\pp,(\pi^0/\gamma_{FSR})\kk\pp$ for mode I or
$\chicj\to\eta\pp\kk$ (no $\eta'$ formed) for mode II.  Also for mode
II, there are small contaminations from the decays
$\psi(3686)\to\gamma\chicj$, $\chicj\to\pi^0\pp\kk$ and
$\chicj\to\gamma\jpsi$ with $\jpsi\to(\gamma/\pi^0)\pp\kk$.  All these
backgrounds have exactly the same topology, or have one less (more)
photon than the signal process, but no $\etap$ intermediate state.
They will produce peaking background in the $\gamma(\gamma)\pp\kk$
invariant mass distribution within the $\chicj$ region.  The
$\gamma(\gamma)\pp\kk$ invariant mass distributions of events with
$\gamma(\gamma)\pp$ mass outside the $\etap$ signal region
($|M(\gamma\pp)-M(\etap)|>15\,\mathrm{MeV}/c^2$,
$|M(\gamma\gamma\pp)-M(\etap)|>25\,\mathrm{MeV}/c^2$) for the two $\etap$
decay modes are shown in Fig.~\ref{2Detapsideband}(b) and
Fig.~\ref{sidebandshape}(b), respectively.  The distributions are
fitted with the sum of three Gaussian functions together with a $3^{rd}$
order polynomial function, which represent the peaking backgrounds and
non-peaking background, respectively. The peaking background shape
obtained here will be used in the following fit as the peaking
background shape within the $\eta'$ signal range.

\section{\boldmath{Signal determination}}
To determine the signal yields, a simultaneous unbinned fit is performed on
the $\gamma (\gamma)\kk\pp$ invariant mass distributions for candidate
events within the $\etap$ signal and sideband regions, where the $\etap$
sideband regions are defined as $25\,\mathrm{MeV}/c^2$
$<|M(\gamma\pp)-M(\etap)|< 40 \mevcc$ and $35\mevcc$
$<M(\GG\pp)-M(\etap)< 85 \mevcc$ for the two $\etap$ decay modes,
respectively.  The following formulas are used to fit the distributions
in the signals and sideband regions, respectively:
\begin{eqnarray}
\begin{split}
& f_{sg}(m) = \sum_{cJ=1}^{cJ=2}N_{cJ}^{sig}\times F_{cJ}^{sig}(m)\otimes G(m,m_i,\sigma_i)\\
            &  + \sum_{i=0}^{i=2}N_{i}^{bkg}\times F_{i}^{bkg}(m)
               + N^{BG}_{signal}\times F^{BG}(m),
\end{split}
\label{equfit1}
\end{eqnarray}
\begin{eqnarray}
\begin{split}
 & f_{sb}(m) = \sum_{i=0}^{i=2}\alpha_i\times N_{i}^{bkg} \times F_{i}^{bkg}(m)~~~~~~~~~~~~~~~~\\
           &     + N^{BG}_{sideband}\times F^{BG}(m),
 \end{split}
\label{equfit2}
\end{eqnarray}
where $F_{cJ}^{sig}(m)$ represents the $\chicj$ signal lineshape,
which is described by the MC simulated shape.  $G(m,m_i,\sigma_i)$ is
a Gaussian function parameterizing the instrumental resolution
difference ($\sigma_i$) and mass offset ($m_i$) between data and MC
simulation, with parameters free in the fit.  Since
$\chi_{c0}\to\etap\kk$ is forbidden by spin-parity conservation, only
the $\chi_{c1,2}$ signals are considered in the fit.  $F_{i}^{bkg}(m)$
is a Gaussian function for peaking backgrounds. MC studies show that the
peaking background shapes do not depend on the $\gamma(\gamma)\pp$
invariant mass. In the fit, the parameters of $F_{i}^{bkg}(m)$ are
identical for $\etap$ signal and sideband regions, and are fixed to
the fitting results from the candidate events with $\gamma(\gamma)\pp$
invariant mass out of the $\etap$ signal region
(Fig.~\ref{2Detapsideband}(b) and Fig.~\ref{sidebandshape}(b)).
$F^{BG}(m)$ represents the non-peaking background which is
parameterized as a $3^{rd}$ order polynomial function.
$N_{cJ}^{sig}$, $N_{i}^{bkg}$, $N^{BG}_{signal}$ and
$N^{BG}_{sideband}$ are the numbers of $\chicj$ signal events, peaking
backgrounds in $\etap$ signal region, and non-peaking background in
$\etap$ signal or sideband region, respectively, to be determined in
the fit.  $\alpha_i$ is the ratio of the number of peaking background
events in the $\etap$ sideband region to that in the $\etap$ signal
region. The magnitudes of $\alpha_i$ are fixed in the fit and the
values are obtained by fitting the $\gamma(\gamma)\pp$ invariant mass
distributions. The detailed procedure to obtain the $\alpha_i$ values is
described in the following.

\begin{figure*}[htbp]
\centering
\begin{overpic}[width=8.0cm,height=5.0cm,angle=0]{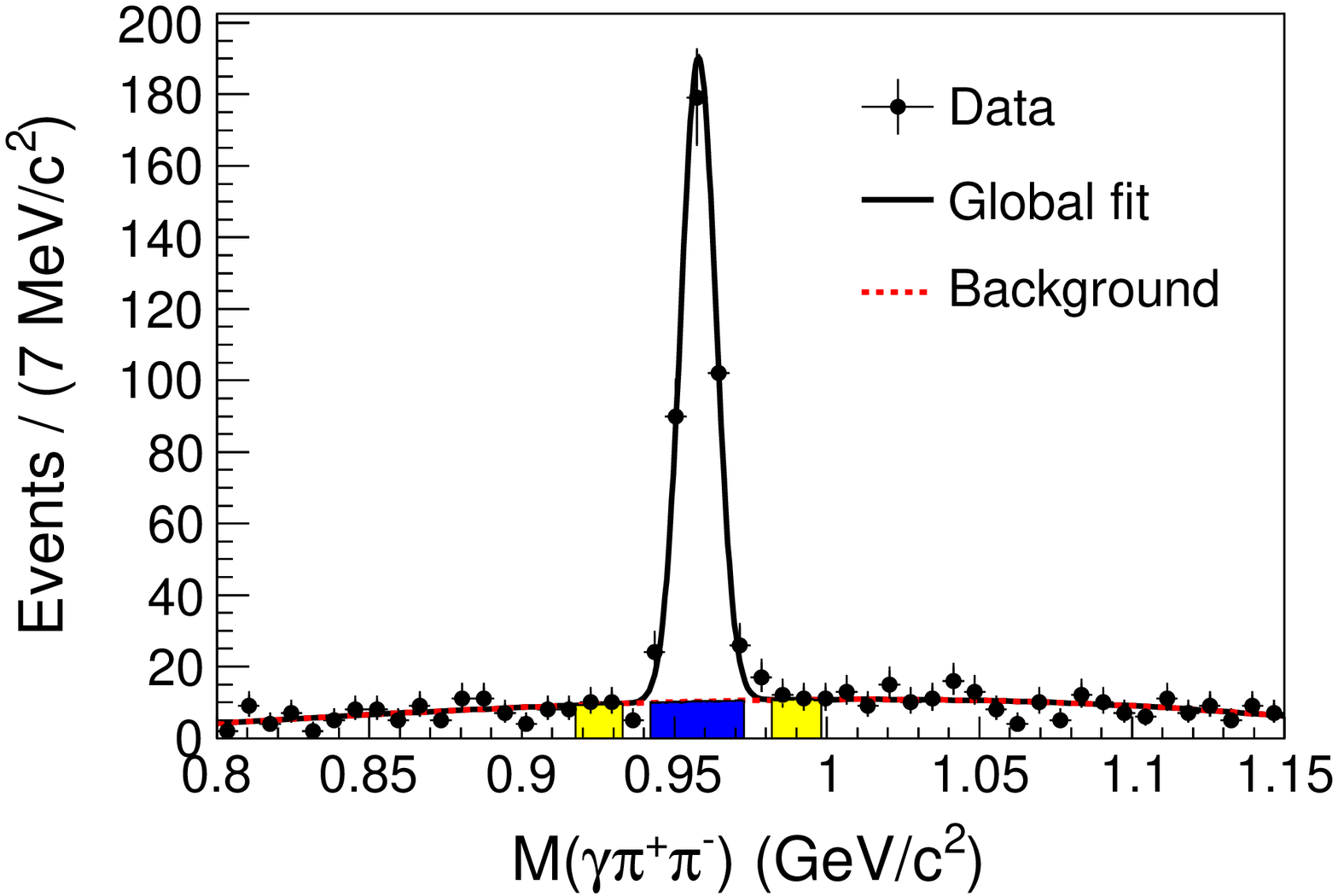}
\put(85,50){\large\bf (a)}
\end{overpic}
\begin{overpic}[width=8.0cm,height=5.0cm,angle=0]{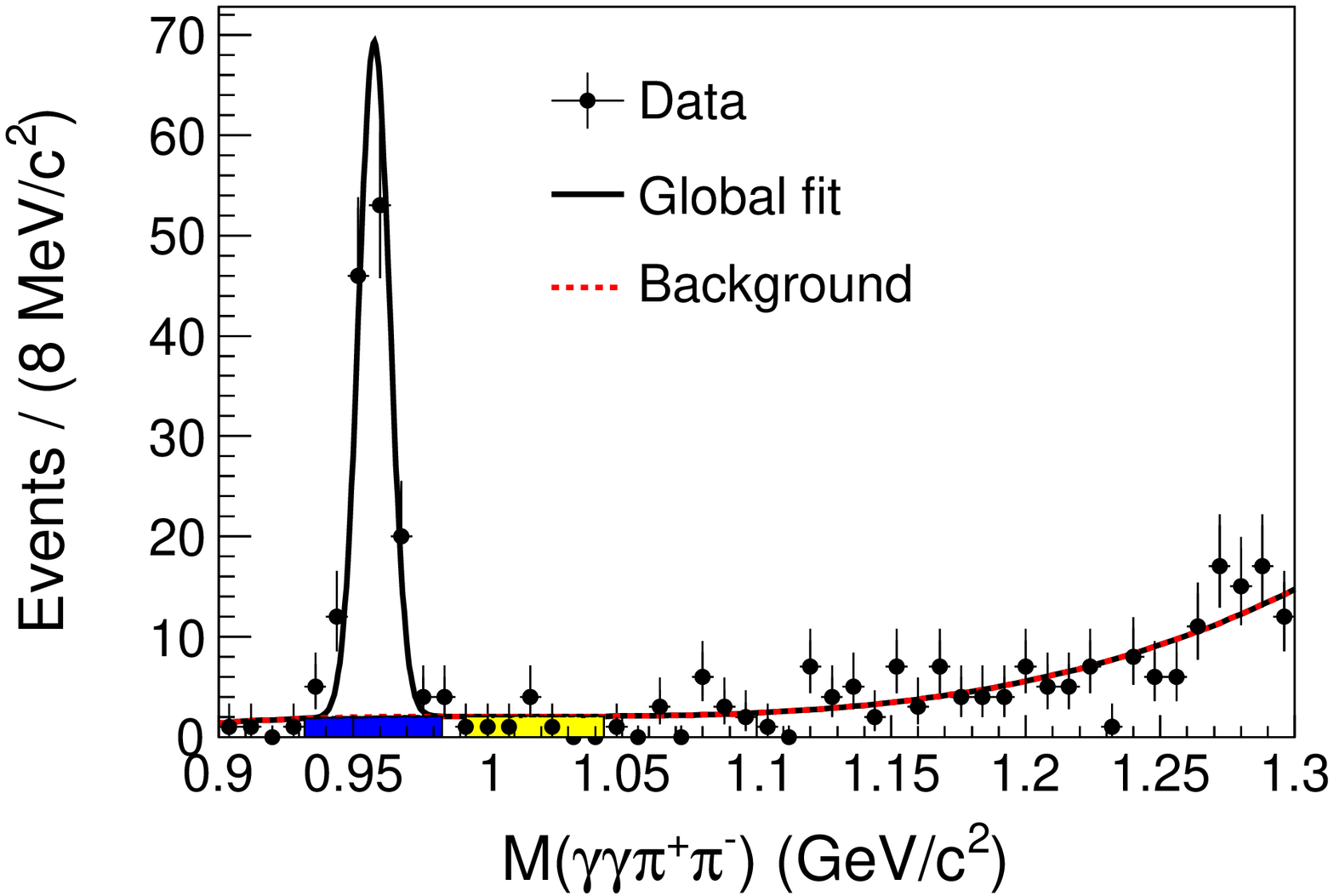}
\put(75,50){\large\bf (b)}
\end{overpic}
\parbox[1cm]{16cm} {\caption{(color online) The $\gamma(\gamma)\pp$ mass distribution
                             within the $\chi_{c1}$ region for (a)
                             $\etap$ decay mode I and (b) $\etap$
                             decay mode II. The band under the peak
                             shows the $\eta'$ signal region, and the
                             other bands show $\eta'$ sideband.}
\label{getratio}
}
\end{figure*}

Figure~\ref{getratio} (a), (b) show the $\gamma(\gamma)\pp$ invariant
 mass distribution for events with $\gamma(\gamma)\pp\kk$ mass within
 the $\chicone$ signal region for the two $\etap$ decay modes,
 respectively. The distributions within $\chiczero$ and $\chictwo$ signal region are similar.
 The $\chicj$ ($J$=0, 1, 2) signal regions are
 defined as
 $|M(\gamma\pp\kk)-M(\chiczero)|<30\mevcc$,
 $|M(\gamma\pp\kk)-M(\chicone)|<15\mevcc$, and
 $|M(\gamma\pp\kk)-M(\chictwo)|<16\mevcc$ for $\eta'$ decay mode
 I, and $|M(\gamma\gamma\pp\kk)-M(\chiczero)|<36\mevcc$,
 $|M(\gamma\gamma\pp\kk)-M(\chicone)|<18\mevcc$, and
 $|M(\gamma\gamma\pp\kk)-M(\chictwo)|<18\mevcc$ for $\eta'$ decay
 mode II.
The distributions are fitted with a Gaussian function which represents
the $\etap$ signal together with a polynomial function which
represents non $\etap$ background.  $\alpha_i$ is the ratio of
integrated polynomial background function in the $\etap$ sideband
region to that in the $\etap$ signal region. Here the background
includes both $\chicj$ peaking background and non-peaking background.
Studies from MC simulation and real data show that the $\chicj$
peaking background and non-peaking background have the same
$\alpha_i$, and the extracted $\alpha_i$ is used in the previous
simultaneous fit.

\begin{figure*}[htbp]
\centering
\begin{overpic}[width=8.0cm,height=5.0cm,angle=0]{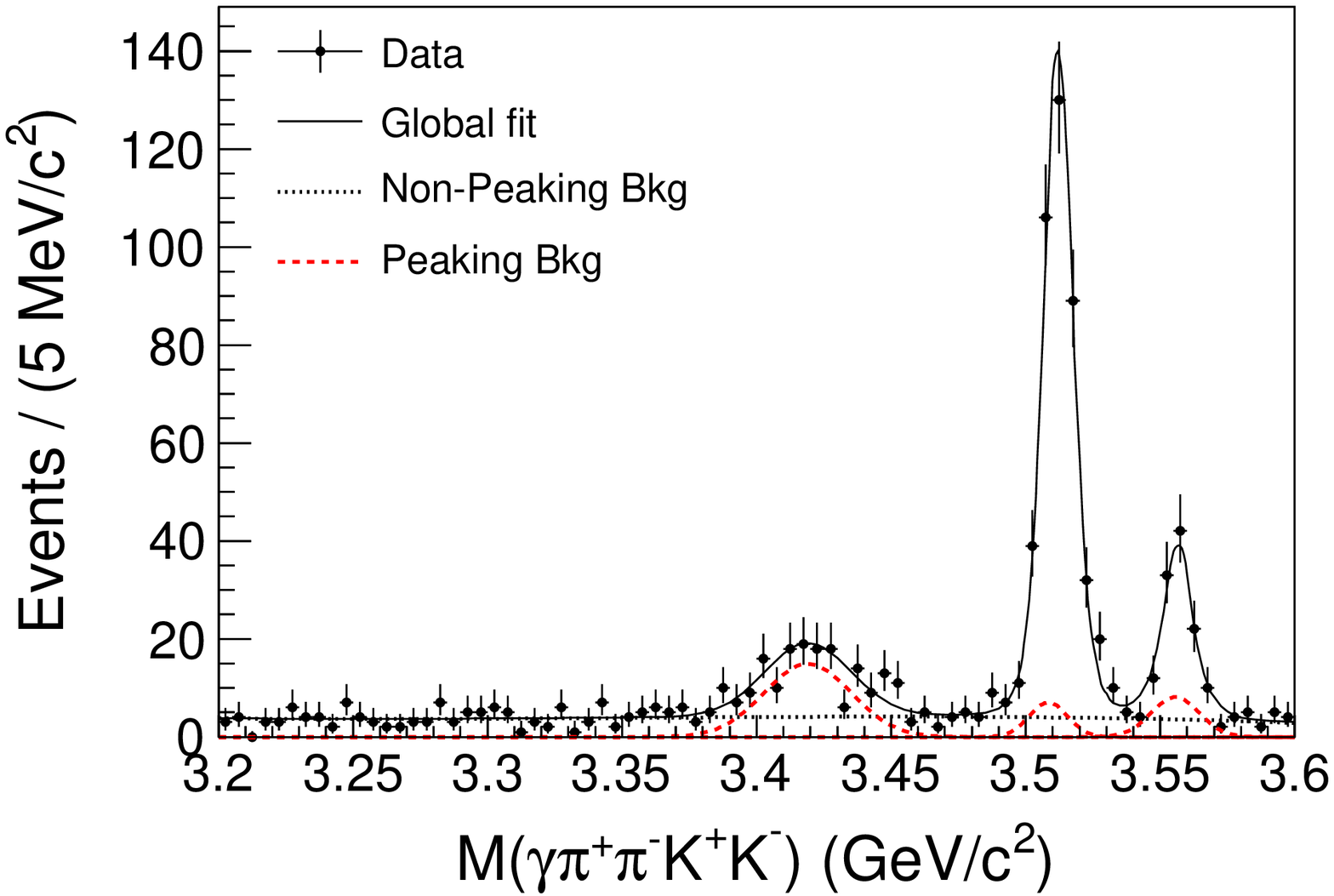}
\put(85,50){\large\bf (a)}
\end{overpic}
\begin{overpic}[width=8.0cm,height=5.0cm,angle=0]{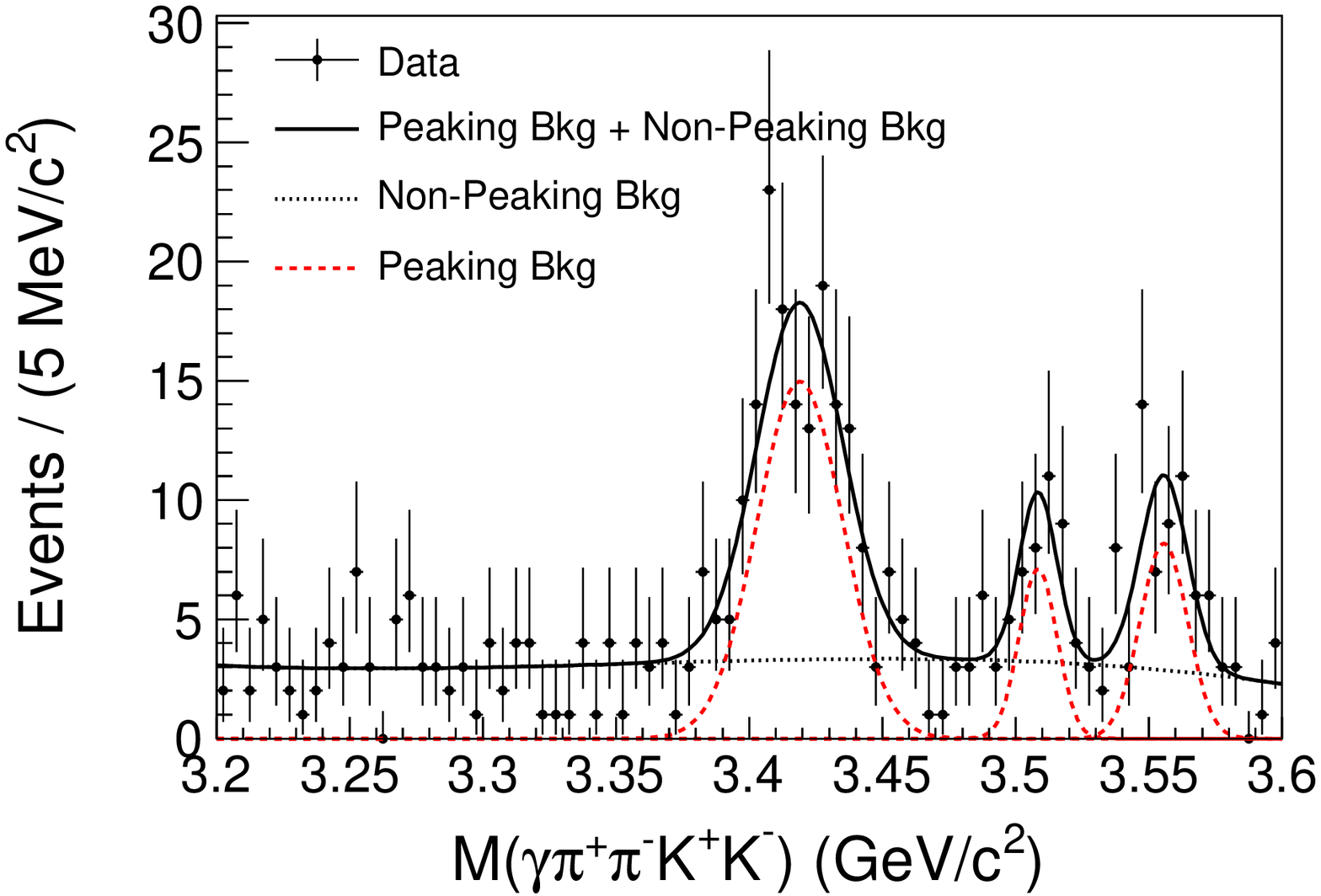}
\put(75,50){\large\bf (b)}
\end{overpic}
\parbox[1cm]{16cm} {\caption{(color online) Invariant mass distribution of
   $\gamma\pp\kk$ for $\etap$ decay mode I in (a) $\etap$ signal
   region and (b) $\etap$ sideband region.}
\label{fitchicj}
}
\end{figure*}

\begin{figure*}[htbp]
\centering
\begin{overpic}[width=8.0cm,height=5.0cm,angle=0]{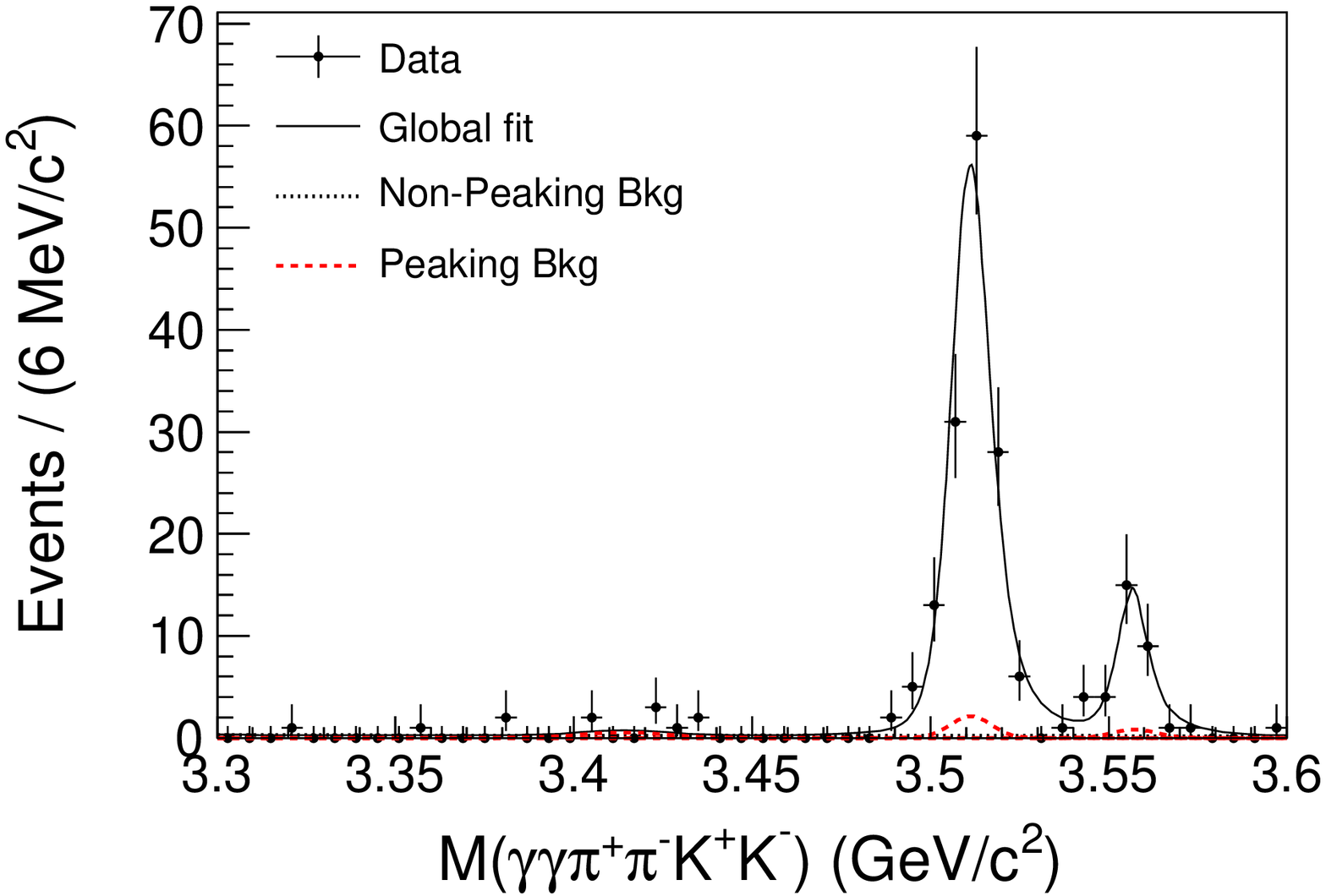}
\put(85,50){\large\bf (a)}
\end{overpic}
\begin{overpic}[width=8.0cm,height=5.0cm,angle=0]{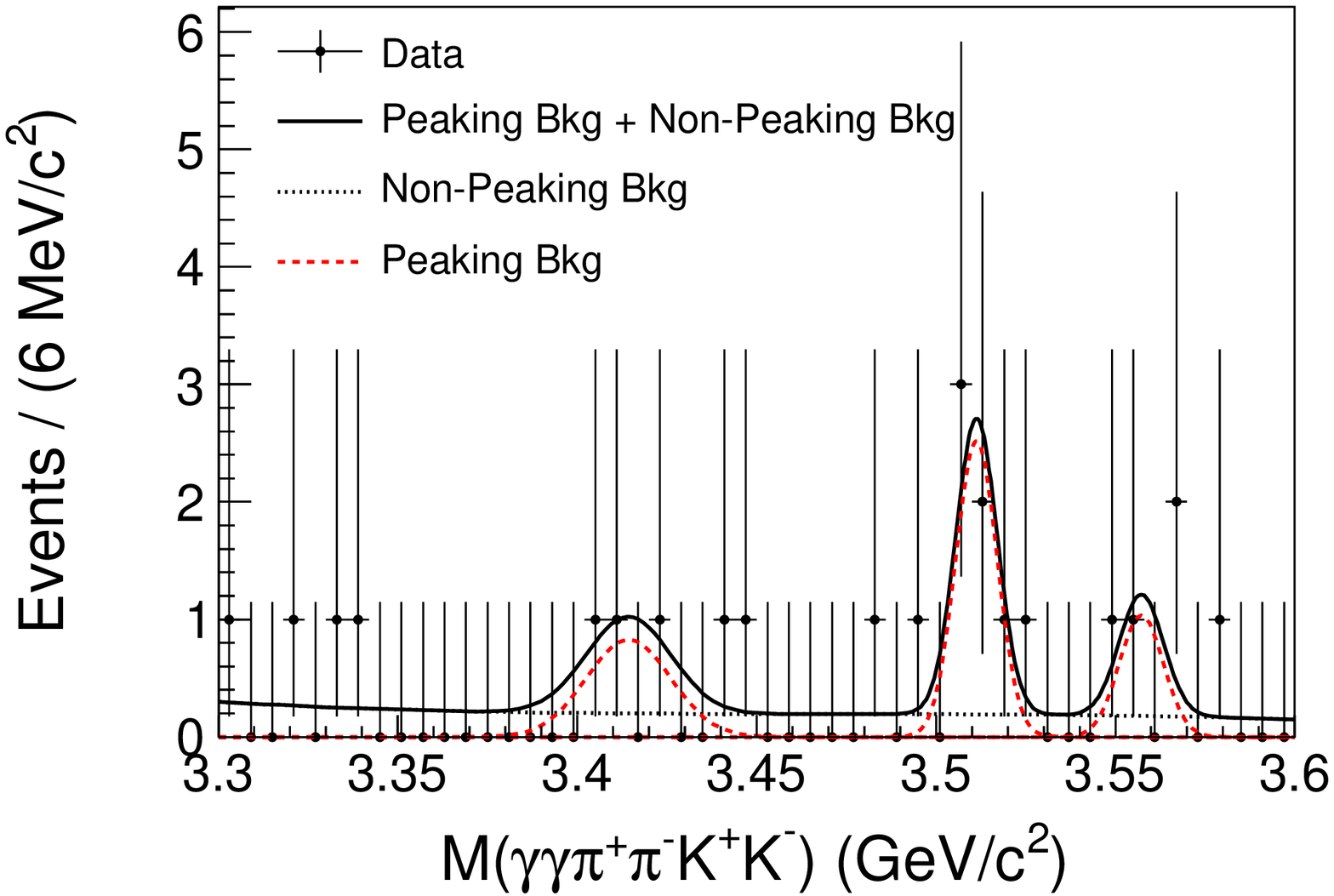}
\put(75,50){\large\bf (b)}
\end{overpic}
\parbox[1cm]{16cm} {\caption{(color online) Invariant mass distribution of
   $\GG\pp\kk$ for the $\etap$ decay mode II in (a) $\etap$ signal
   region and (b) $\etap$ sideband region. The fraction of non-peaking background is very small so its line is invisible in left plot.}
\label{fitchicj2}
}
\end{figure*}

The $\gamma(\gamma)\pp\kk$ invariant mass distributions of candidate
events in $\etap$ signal and sideband regions for the two $\etap$
decay modes are shown in Figs.~\ref{fitchicj} and \ref{fitchicj2},
respectively.  The simultaneous unbined fits are carried out to
determine the signal yields, and the results are summarized in
Table~\ref{Brhaha}.

\section{\boldmath{Branching fraction}}
The branching fractions of $\chicj\to\etapkk$ in the two $\etap$ decay modes are
calculated according to:
\begin {equation}
\label{eqbr1}
\begin{split}
&\mathcal{B}_1(\chicj\to\etapkk)= \\
&\frac {N_{cJ}^{sig}}{N_{\psi(3686)}\times \mathcal{B}(\psi(3686)\to\gamma\chicj)
                                                       \times \mathcal{B}(\etaptogammarho)
                                                       \times \epsilon^1_{cJ}}
~~~~~~~~~~~~~~~~~~~~~~
\end{split}
\end {equation}
\begin {equation}
\label{eqbr2}
\begin{split}
&\mathcal{B}_2(\chicj\to\etapkk)=\\
&\quad\frac {N_{cJ}^{sig}}{N_{\psi(3686)}\times \mathcal{B}(\psi(3686)\to\gamma\chicj)}~~~~~~~~~~~~~~~~~~~~~~~~~~~~\\
&\quad\times\frac{1}{                                              \mathcal{B}(\etaptoetapipi)
                                                    \times \mathcal{B}(\eta\to\GG)
                                                    \times \epsilon^2_{cJ}}
\end{split}
\end {equation}
where $N_{cJ}^{sig}$ is the number of signal events extracted from the
simultaneous unbinned fit.  $N_{\psi(3686)}$ is the number of
$\psi(3686)$ events.  $\mathcal{B}(\psi(3686)\to\gamma\chicj)$,
$\mathcal{B}(\etaptogammarho)$, $\mathcal{B}(\etaptoetapipi)$ and
$\mathcal{B}(\eta\to\GG)$ are branching fractions from the
PDG~\cite{pdg}. $\epsilon^1_{cJ}$ and $\epsilon^2_{cJ}$ are the detection
efficiencies for mode I and mode II, respectively.  Detailed studies
in Sec.~\ref{PWA} show that abundant structures are observed in the
$\kk$ and $\etap K^\pm$ invariant mass spectra. To get the detection
efficiencies properly, a partial wave analysis (PWA) using covariant
tensor amplitudes is performed on the candidate events, and the
detection efficiencies are obtained from MC samples generated with the
differential cross section from the PWA results. The detection
efficiencies and the branching fractions (statistical uncertainty
only) are also shown in Table~\ref{Brhaha}.
\begin{table*}[htbp]
\centering
\caption{Summary for the fit results, detection efficiencies and branching fractions (statistical uncertainty only).
\label{Brhaha}
}
\begin{tabular}{clccccc} \hline \hline
                              &                  &~~~~$N^{sig}_{cJ}$~~~~&~~~~~$N^{bkg}_{i}$~~~~~&$\alpha_i$&~~$\epsilon$(\%)~~ &~~$\mathcal{B}(\chicj\to\etapkk)$($10^{-4}$)\\ \hline
~~~~~\multirow{2}*{$\chiczero$ }~~~~~   & $\etaptogammarho$& $\cdot\cdot\cdot$      & $121\pm11$ &$0.977\pm0.002$  & $\cdot\cdot\cdot$  &  $\cdot\cdot\cdot$  \\
\multirow{2}*{}               & $\etaptoetapipi$ &$\cdot\cdot\cdot$ & $3\pm2$     &$ 1.7\pm0.3$& $\cdot\cdot\cdot$    & $\cdot\cdot\cdot$                     \\  \hline
\multirow{2}*{$\chi_{c1}$}    & $\etaptogammarho$& $388\pm23$     & $25\pm7$ &$0.984\pm0.004$   &14.88       & $9.09\pm0.54$                    \\
\multirow{2}*{}               & $\etaptoetapipi$ & $141\pm13$     & $5\pm2$     &$1.3\pm0.3$ &10.14       & $8.33\pm0.77$                    \\ \hline
\multirow{2}*{$\chi_{c2}$}    & $\etaptogammarho$& $77\pm13$      & $36\pm8$  &$0.979\pm0.003$   &15.38       & $1.84\pm0.31$                    \\
\multirow{2}*{}               & $\etaptoetapipi$ & $30\pm6 $      & $2\pm2$     &$1.4\pm0.4$ &9.25        & $2.05\pm0.41$                    \\ \hline\hline
\end{tabular}
\end{table*}

\section{\boldmath{Estimation of Systematic Uncertainties}}
Several sources of systematic uncertainties are considered in the
measurement of branching fractions. These include the differences
between data and MC simulation for the tracking, PID, photon
detection, kinematic fit, fitting procedure and number of $\psi(3686)$
events as well as the uncertainties in intermediate resonance decay
branching fractions.

\emph{a. Tracking and PID~~} The uncertainties from tracking and PID
efficiency of the kaon are investigated using an almost background free
control sample of $\jpsi\to K_S^0K^{\pm}\pi^{\mp}$ from
$(225.2\pm2.8)\times 10^6$ $\jpsi$ decays~\cite{jpsinumber}. Both kaon
tracking efficiency and PID efficiency are studied as a function of
transverse momentum and polar angle. The data-MC simulation
differences are estimated to be 1\% per track for the tracking efficiency
and 2\%~\cite{track} per track for the PID efficiency.  Therefore, 2\%
uncertainty for the tracking efficiency and 4\% uncertainty for the PID
efficiency are taken as the systematic uncertainties for two kaons.  The
uncertainty for the pion tracking is investigated with high
statistics, low background samples of $\jpsi\to\rho\pi$, $\jpsi\to
p\bar p\pp$ and $\psi(3686)\to\pp\jpsi$ with $\jpsi\to l^+l^-$
events. The systematic uncertainty is taken to be 1\% per
track~\cite{pitrack}, and 2\% for two pions.

\emph{b. Photon detection efficiency~~} The uncertainty due to photon
detection and reconstruction is 1\% per photon~\cite{track}.  This
value is determined from studies using clean control samples, such as
$\jpsi\to\rho^0\pi^0$ and $e^+e^-\to\gamma\gamma$. Therefore,
uncertainties of 2\% and 3\% are taken for photon detection
efficiencies in the two $\etap$ decay modes, respectively.

\emph{c. Kinematic fit~~} To investigate the systematic uncertainty
from the 4C kinematic fit, a clean control sample of $\jpsi\to\eta\phi$,
$\eta\to\pp\pi^0$, $\phi\to\kk$, which has a similar final state to those
of this analysis, is selected.  A 4C kinematic fit is applied
to the control sample, and the corresponding efficiency is estimated from
the ratio of the number of events with and without the kinematic fit. The
difference of efficiency between data and MC simulation, 3.3\%, is
taken as the systematic uncertainty.

\emph{d. Mass window requirements~~} Several mass window requirements
are applied in the analysis. In mode I, mass windows on
$M(\GG)_{rec}$ and $M(\pp)_{rec}$ are applied to suppress
backgrounds with $\jpsi$ intermediate states, $M(\GG)$ requirements
are used to remove backgrounds with $\pi^0$ in the final state,
and an $M(\gamma\pp)$ requirement is used to determine the $\etap$
signal. In mode II, mass windows on $M(\GG)$ are used to remove
backgrounds with $\pi^0$ and to determine the $\eta$ signal. An
$M(\GG\pp)$ mass window is used for the $\etap$ signal.  Different
values of these mass window requirements within $3\sigma\sim5\sigma$
($\sigma$ is the corresponding mass resolution) have been used, and
the largest differences in the branching fractions are taken as
systematic uncertainties.

\emph{e. Fitting procedure~~} As described above, the yields of the
$\chicj$ signal events are derived from the simultaneous unbinned fits
to the invariant mass of $\gamma(\gamma)\kk\pp$ with $\gamma(\gamma)
\pp$ invariant mass within the $\etap$ signal and sideband regions for the
two $\etap$ decay modes, respectively. To evaluate the systematic
uncertainty associated with the fitting procedure, the following
aspects have been studied.  {\it 1) shape of non-peaking background:}
The uncertainties due to the non-peaking background parameterization
are estimated by the difference when we use a $2^{nd}$ or
$4^{th}$ instead of a $3^{rd}$ order background polynomial function.
{\it 2) shape of peaking backgrounds:} In the nominal fit, shapes of
peaking backgrounds are fixed to the fitting results of events with
$\gamma(\gamma)\pp$ mass outside the $\etap$ signal region
(Fig.~\ref{2Detapsideband}(b), Fig.~\ref{sidebandshape}(b)).
Alternative shapes of peaking background obtained from different
$\gamma(\gamma)\pp$ regions are used to constrain the shape of peaking
background in the fit, and to estimate the corresponding systematic uncertainty.
{\it 3) fitting range:} A series of fits with different intervals on the $\gamma(\gamma)\kk\pp$
invariant mass spectrum are performed.  {\it 4) sideband range:} The candidate
events with $\gamma(\gamma)\pp$ invariant mass within the $\etap$
sideband region are used to constrain the amplitude of peaking
backgrounds in the fits. The corresponding systematic uncertainties are estimated
with different interval of sideband ranges with width from $1\sigma_{\eta'}$ to
$3\sigma_{\eta'}$ ($\sigma_{\eta'}$ is the width of the nominal sideband range).
 {\it 5) the normalization
factor:} The normalization factors $\alpha_i$ are varied within their
uncertainties listed in Table~\ref{Brhaha}.  The systematic
uncertainties of these aspects are taken as the largest
differences in the branching fractions to the nominal result.

\emph{f. Detection efficiency~~} As mentioned previously, abundant
structures are observed in both $\kk$ and $\eta K^\pm$ invariant mass
spectra, respectively.  A full PWA is performed to estimate the
detection efficiencies of the $\chicone$ signal, and the following two
aspects are considered to evaluate the detection efficiency
uncertainties: {\it 1)} The statistical uncertainties of PWA fit
parameters (the magnitudes and phases of partial waves), which are
obtained from the PWA results; {\it 2)} The uncertainties of input
mass and width of intermediate states~\cite{pdg}.  For the $\chictwo$
signal, a simple PWA is performed on the candidate events,
and the detection efficiency uncertainties are estimated by the
differences of PWA fitting with or without background subtraction.

\emph{g. Other systematic uncertainties~~} The number of $\psi(3686)$
events is determined from an inclusive analysis of $\psi(3686)$
hadronic events with an uncertainty of 0.8\%~\cite{totaln}. The
uncertainties due to the branching fractions of
$\psi(3686)\to\gamma\chicj$, $\etap\to\gamma\rho^0$, $\etap\to\eta\pp$
and $\eta\to\gamma\gamma$ are taken from PDG~\cite{pdg}.

A summary of all the uncertainties is shown in
Table~\ref{SysErrBr1}. The total systematic uncertainty is obtained by
summing all individual contributions in quadrature.

The final branching fractions of $\chi_{c1,2}\to\etap\kk$ measured
from the two $\etap$ decay modes are listed in
Table~\ref{lastbrresult}, where the first uncertainties are
statistical, and second ones are systematic.  The measured branching
fractions from the two $\etap$ decay modes are consistent with each
other within their uncertainties. The measurements from the two decay
modes are, therefore, combined by considering the correlation of
uncertainties between the two measurements, the mean value and the
uncertainty are calculated with~\cite{TPLi},
\begin {equation}
\label{eq11}
\overline{x}\pm\sigma(\overline{x})=\frac{\sum_{j}(x_j\cdot\sum_{i}\omega_{ij})}
                                        {\sum_i\sum_j\omega_{ij}}
                                \pm\sqrt{\frac{1}{\sum_i\sum_j\omega_{ij}}}~~,\\
\end {equation}
where $i$ and $j$ are summed over all decay modes, $\omega_{ij}$ is
the element of the weight matrix $W=V^{-1}_{x}$, and $V_{x}$ is the
covariance error matrix calculated according to the statistical
uncertainties listed in Table~\ref{Brhaha} and the systematic
uncertainties listed in Table~\ref{SysErrBr1}. When combining the
results of the two decay modes, the error matrix can be calculated as
\begin{equation}
V=
\left(
 \begin{array}{cc}
 \sigma_{1}^2+\epsilon_f^2x_1^2 & \epsilon_f^2x_1x_2   \\
 \epsilon_f^2x_1x_2             & \sigma_2^2+\epsilon_f^2x_2^2\\
 \end{array}
\right),
\end{equation}
where $\sigma_i$ is the independent absolute uncertainty (the
statistical uncertainty and all independent systematical uncertainties
added in quadrature) in the measurement mode $i$, and $\epsilon_f$ is
the common relative systematic uncertainties between the two
measurements (All the common systematic uncertainties added in
quadrature. The items in Table.~\ref{SysErrBr1} with $' * '$ are
common uncertainties, and the other items are independent
uncertainties). $x_i$ is the measured value given by mode $i$.  Then
the combined mean value and combined uncertainty can be calculated as
:
\begin{equation}
\overline{x}=\frac{x_1\sigma_2^2+x_2\sigma_1^2}
        {\sigma_1^2+\sigma_2^2+(x_1-x_2)^2\epsilon_f^2} .
\label{com1}
\end{equation}
\begin{equation}
\sigma^2(\overline{x})=\frac{\sigma_1^2\sigma_2^2+(x_1^2\sigma_2^2+x_2^2\sigma_1^2)\epsilon_f^2}
                       {\sigma_1^2+\sigma_2^2+(x_1-x_2)^2\epsilon_f^2} .
\label{com2}
\end {equation}
 The calculated results are shown in Table~\ref{lastbrresult}.

\begin {table*}[htp]
\begin {center}
\caption {Summary of systematic uncertainties (in \%) for the branching fractions $\chi_{c1,2}\to\etap\kk$. The items with $' * '$ are common uncertainties of two $\eta'$ decay modes.
\label{SysErrBr1}
}
\begin {tabular}{lcccc} \hline \hline
           & \multicolumn{2}{c}{$\etap\to\gamma\rho^0$} & \multicolumn{2}{c}{$\etap\to\eta\pp$} \\
Source     & ~~~$\chicone(\%)$~~~  &  ~~~$\chictwo(\%)~~~$  &  ~~~$\chicone(\%)$~~~  & ~~~$\chictwo$(\%)~~~ \\ \hline
*Tracking efficiency            & 4.0 & 4.0 & 4.0  & 4.0  \\
*Particle identification         & 4.0 & 4.0 & 4.0  & 4.0  \\
*Photon detection efficiency     & 2.0 & 2.0 & 3.0  & 3.0  \\
4C kinematic fit                 & 3.3 & 3.3 & 3.3  & 3.3  \\
Mass windows                     & 0.8 &12.5 & 2.6  & 3.9  \\
Non-peaking background shape     & 1.6 & 0.0 & 0.7  & 3.0  \\
Peaking background shape         & 3.4 & 5.2 & 1.0  & 0.0  \\
Fit range                        & 2.2 & 2.7 & 0.7  & 3.0  \\
Sideband range                   & 0.2 & 7.6 & 0.7  & 3.0  \\
Normalization factor             &0.0  &0.1  &1.1   & 3.3\\
Efficiency                       & 0.4 & 2.7 & 0.7  & 4.6  \\
*Number of $\psi(3686)$ events              & 0.8 & 0.8 & 0.8  & 0.8  \\
*$\mathcal{B}(\psi(3686)\to\gamma\chicj)$   & 4.3 & 3.9 & 4.3  & 3.9  \\
$\mathcal{B}(\etap\to\gamma\rho^0/\eta\pp)$& 2.0 & 2.0 & 1.6  & 1.6  \\
$\mathcal{B}(\eta\to\gamma\gamma)$         & -   & -   & 0.5  & 0.5  \\ \hline
Total                            &9.5  &18.0 & 9.2  & 12.0 \\ \hline \hline
\end {tabular}
\end {center}
\end {table*}

\section{\boldmath{Partial Wave Analysis Of} $\chi_{c1}\to\etapkk$}
\begin{figure*}[htbp]
\centering
\begin{overpic}[width=0.9\linewidth,height=0.35\linewidth,angle=0]{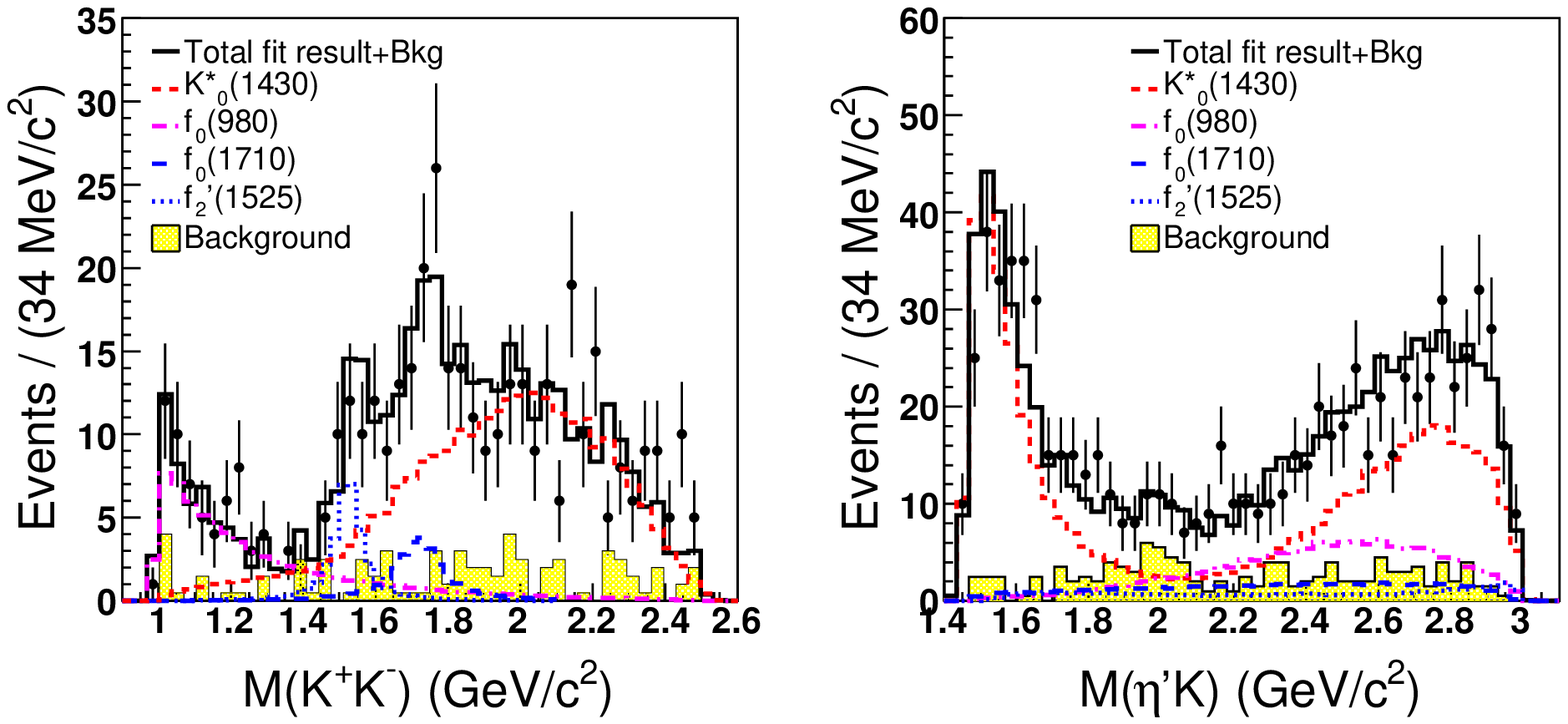}
\put(42,32){\large\bf (a)}
\put(92,32){\large\bf (b)}
\end{overpic}
\begin{overpic}[width=0.9\linewidth,height=0.35\linewidth,angle=0]{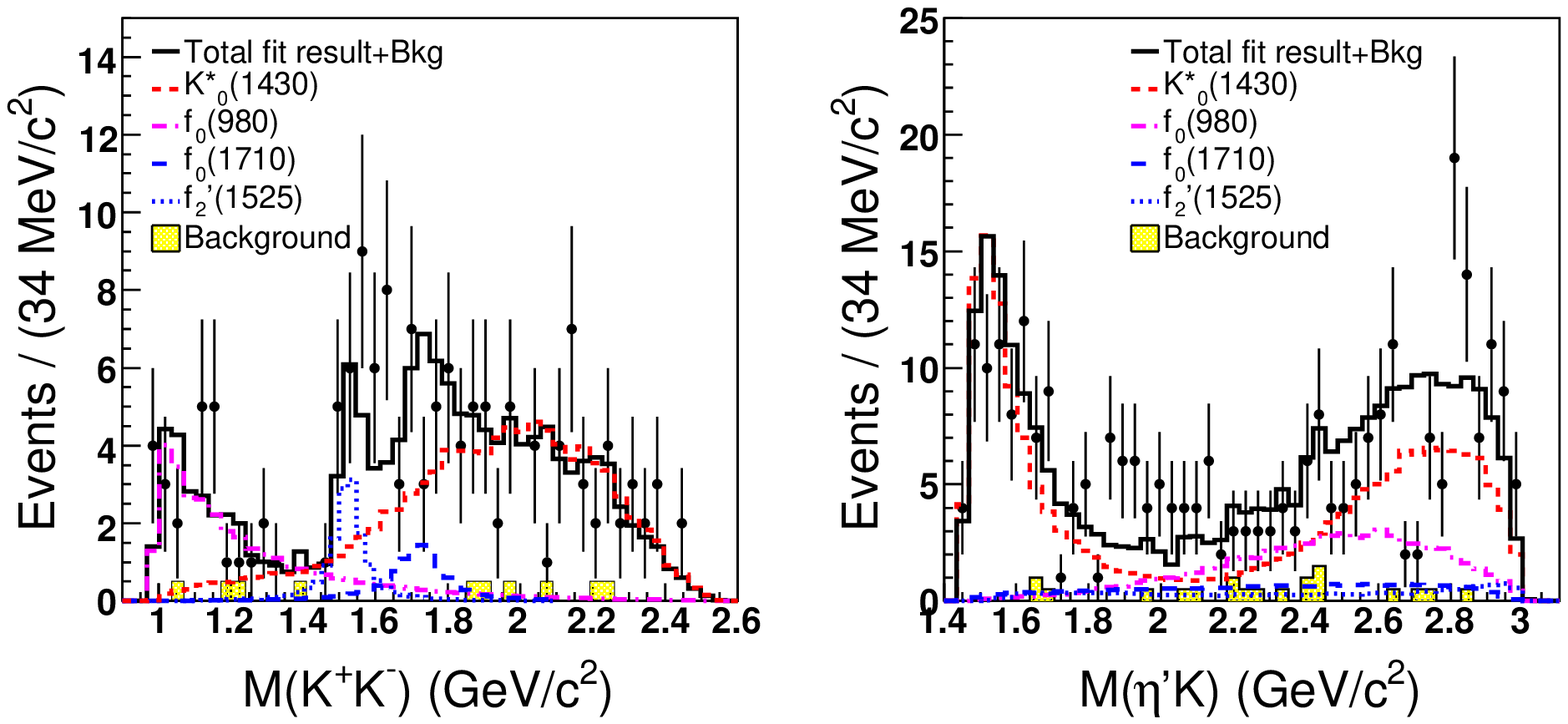}
\put(42,32){\large\bf (c)}
\put(92,32){\large\bf (d)}
\end{overpic}
\parbox[1cm]{16cm} {
\caption{(color online) The invariant mass distributions of $\kk$ and $\eta' K^\pm$
  within the $\chicone$ mass range.
(a)(b) for the $\etap$ decay mode I, and (c)(d) for the $\etap$ decay mode II.}
\label{project}
}
\end{figure*}

{\label{PWA}}
As shown in Fig.~\ref{project}, there are abundant structures observed
in the $\kk$ and $\etap K^\pm$ invariant mass distributions.  In the $\kk$
invariant mass spectrum, an $f_0(980)$ is observed at $\kk$
threshold. There are also structures observed around $1.5\gevcc$ and
$1.7\gevcc$.  In the $\etap K^\pm$ invariant mass spectrum, a
structure is observed at threshold, which might be a
$K_0^{\ast\pm}(1430)$ or other excited kaon with different
$J^{P}$ at around $1.4\gevcc$.  To study the sub-processes with
different intermediate states and to evaluate the detection
efficiencies of the decay $\chicj\to\etap\kk$ properly, a PWA is
performed on $\chicj$ signal candidates with the combined data of
the two $\etap$ decay modes.
\subsection{\boldmath{Decay amplitude and likelihood construction}}{\label{bwform}}
In the PWA, the sub-processes with following sequential two-body decays are
considered:
\begin{itemize}
\item[1.] $\psi(3686)\to\gamma+\chicone$, $\chicone\to\etap + f_0(X)/f_2(X)$,\\
                                       $f_0(X)/f_2(X)\to \kk$;
\item[2.] $\psi(3686) \to \gamma + \chicone$, ~~$\chicone \to K^{\ast\pm}_X + K^{\mp}$,
                                       ~~$K^{\ast\pm}_X \to \etap K^{\pm}$;
\end{itemize}
The  2-body decay amplitudes are constructed in the covariant tensor
formalism~\cite{zoubs}, and the radius of the centrifugal barrier is set to be
$1.0$ fm.  Due to limited statistics in the fit, the lineshape of
intermediate states, e.g.  $f_0(980)$, $f_0(1710)$, $f_2'(1525)$ and
$K_X^{\ast\pm}(1430)$ etc, are all taken from the literature and fixed in the fit.  The shape
of $f_0(980)$ is described with the Flatt\'{e} formula~\cite{Flatte}:
\begin{equation}
\dfrac{1}{M^2-s-i(g_1\rho_{\pi\pi}+g_2\rho_{KK})},
\label{Flatte}
\end{equation}
where $s$ is the $\kk$ invariant mass-squared, and $\rho_{\pi\pi}$ and
$\rho_{KK}$ are Lorentz invariant phase space factors, $g_{1,2}$ are
coupling constants to the corresponding final state, and the parameters
are fixed to values measured in \hbox{BESII}~\cite{bes2980}:
$M=0.965\gevcc$,
$g_1=0.165\,\mathrm{GeV}^2/c^4$, and $g_2/g_1=4.21$.  The $f_2'(1525)$
and $f_0(1710)$ are parameterized with the Breit-Wigner propagator
with constant width:
\begin{equation}
BW(s)=\frac{1}{M^{2}_{R}-s-iM_{R}\Gamma_{R}},
\end{equation}
where $M_R$ and $\Gamma_R$ are the mass and width of the resonances,
respectively, and are fixed at PDG values~\cite{pdg}.  The excited
kaon states at the $\etap K^\pm$ invariant mass threshold are
parameterized with the Flatt\'{e} formula:
\begin{equation}
\dfrac{1}{M^2-s-i(g_1\rho_{K\pi}(s)+g_2\rho_{\eta'K}(s))},
\label{cleoflatte}
\end{equation}
where $s$ is the $\eta'K$ invariant mass-squared, $\rho_{K\pi}$ and
$\rho_{\etap K}$ are Lorentz invariant phase space factors, $g_{1,2}$
are coupling constants to the corresponding final state.  The
parameters of $K_0^{\ast\pm}(1430)$ are fixed to values measured by
CLEO~\cite{cleokstar}:  $M=1.4712\gevcc$, $g_1=0.2990
\,\mathrm{GeV}^2/c^4$, and $g_2=0.0529\,\mathrm{GeV}^2/c^4$.

The decay  amplitude is constructed as follows~\cite{zoubs} :
\begin {eqnarray}
\begin{split}
A=&\psi_{\mu}(m_1)e^*_{\nu}(m_2)A^{\mu\nu}\\
=&\psi_{\mu}(m_1)e^*_{\nu}(m_2)\sum_i^{j=1,2}\Lambda_{ij}U^{\mu\nu}_{ij} ,
\end{split}
\label{amp1}
\end {eqnarray}
\begin{equation}
\Lambda_{ij}=\rho_{ij} e^{i\phi_{ij}} ~~(j=1,2, \phi_{i1}=\phi_{i2}),
\label{amp2}
\end{equation}
\begin{equation}
U^{\mu\nu}_{ij}= BW_{\chi_{cJ}}\times BW_i\times A_{ij}(J^{PC}) ,
\label{amp3}
\end{equation}
where $\psi_{\mu}(m_1)$ is the polarization vector of $\psi(3686)$,
$e_{\nu}(m_2)$ is the photon polarization vector, and
$U^{\mu\nu}_{ij}$ is the amplitude of the $i$th state. For
$\psi(3686)\to\gamma+\chi_{c1},\chi_{c1}\to \eta'+X_i ~/~K^{\pm}+
X_i$, each intermediate state $X_i$ will introduce two independent
amplitudes, which are identified by the subscript $j=1,2$. The
detailed formulas for $U^{\mu\nu}_{ij}$ for states with different
$J^{PC}$, which are the same as those for
$\psi\to\gamma\eta\pi^+\pi^-$, can be found in reference~\cite{zoubs}.
$\rho_{ij}$ is the magnitude and $\phi_{ij}$ is the phase angle of the
amplitude of the $i$-$th$ state. In the fit, the phase of the two
amplitudes of the same states are set to be same,
$\phi_{i1}=\phi_{i2}$. $BW_{\chi_{cJ}}$ and $BW_i$ are the propagators
for $\chicj$ and the intermediate states observed in the $K^+K^-$ or
$\etap K^\pm$ invariant mass spectra, respectively.  $A_{ij}(J^{PC})$
is the remaining part that is dependent on the $J^{PC}$ of the
intermediate states.  Since all the parameters in the propagators are
fixed in the fit, there are three free parameters (two magnitudes and
one phase) for each state in the fit. The total differential cross
section $d\sigma/d\phi$ is
\begin{equation}
\begin{split}
&\frac{d\sigma}{d\phi}=\frac{1}{2}\times\\
&\sum_{m_1=1}^2\sum_{m_2=1}^2\psi_{\mu}(m_1)e^*_{\nu}(m_2)A^{\mu\nu}\psi^*_{\mu'}(m_1)e_{\nu'}(m_2)A^{*\mu'\nu'}.
\end{split}
\label{difsigma}
\end{equation}

The relative magnitudes and phases of each sub-process are determined
by an unbinned maximum likelihood fit.
The probability to observe the event characterized by the measurement
$\xi_i$ is the differential cross section normalized to unity:
\begin{equation}
P(\xi_i,\alpha)=\frac{\omega(\xi_i,\alpha)\epsilon(\xi_i)}
{\int d\xi_i\omega(\xi_i, \alpha)\epsilon(\xi_i)},
\end{equation}
where $\omega(\xi_i, \alpha)\equiv(\frac{d\sigma}{d\phi})_i$, $\alpha$
is a set of unknown parameters to be determined in the fitting, and
$\epsilon(\xi_i)$ is the detection efficiency. The joint probability
density for observing $N$ events in the data sample is:
\begin{equation}
 \mathcal{L} = \prod^{N}_{i=1} P(\xi_{i},\alpha) = \prod^{N}_{i=1}
\frac{\omega(\xi_i,\alpha)\epsilon(\xi_i)}{\int d\xi_i\omega(\xi_i, \alpha)\epsilon(\xi_i)} .
\end{equation}
FUMILI~\cite{fumili} is used to optimize the fit parameters to achieve the
maximum likelihood value. Technically, rather than maximizing $\mathcal{L}$,
$\mathcal{S} = - \ln\mathcal{L}$ is minimized, i.e.,
\begin{equation}
\mathcal{S} = -\ln\mathcal{L}
= -\sum^{N}_{i=1} \ln(\frac{\omega(\xi_i,\alpha)}
  {\int d\xi_i\omega(\xi_i,\alpha)\epsilon(\xi_i)})
   - \sum^{N}_{i=1}\ln\epsilon(\xi_i).
\end{equation}
For a given data set, the second term is a constant and has no impact
on the relative changes of the $\mathcal{S}$ value.  In practice, the
normalized integral $\int d\xi_i\omega(\xi_i,\alpha)\epsilon(\xi_i)$
is evaluated by the PHSP MC samples. The details of the PWA fit
process are described in Ref.~\cite{pwamethod}.

\subsection{\boldmath{Background treatment}}

In this analysis, background contamination in the signal region is
estimated from events within different sideband regions.  The $\etap$
signal region is defined with the requirement (I)
$|M(\gamma\pp)-M(\etap)|<15 \mevcc$ for mode I, or
$|M(\gamma\gamma\pp)-M(\etap)|<25\mevcc$ for mode II.  While the
$\etap$ sideband region is defined with the requirement (II) $20
\mevcc<|M(\gamma\pp)-M(\etap)|<50\mevcc$ or
$30\mevcc<|M(\gamma\gamma\pp)-M(\etap)|<80 \mevcc$, respectively.
The $\chicone$ signal region is defined with the requirement (III)
$|M(\gamma\pp\kk)-M(\chicone)|<15\mevcc$ or
$|M(\gamma\gamma\pp\kk)-M(\chicone)|<18\mevcc$ for the two $\etap$
decay modes, respectively.  The $\chicone$ sideband region is defined
with requirement (IV) $20 \mevcc<M(\chicone)-M(\gamma\pp\kk)<50
\mevcc$ or $23\mevcc<M(\chicone)-M(\gamma\gamma\pp\kk)<59\mevcc$
for modes I and II, respectively.
\begin{figure*}[htbp]
\centering
\begin{overpic}[width=8.0cm,height=5.0cm,angle=0]{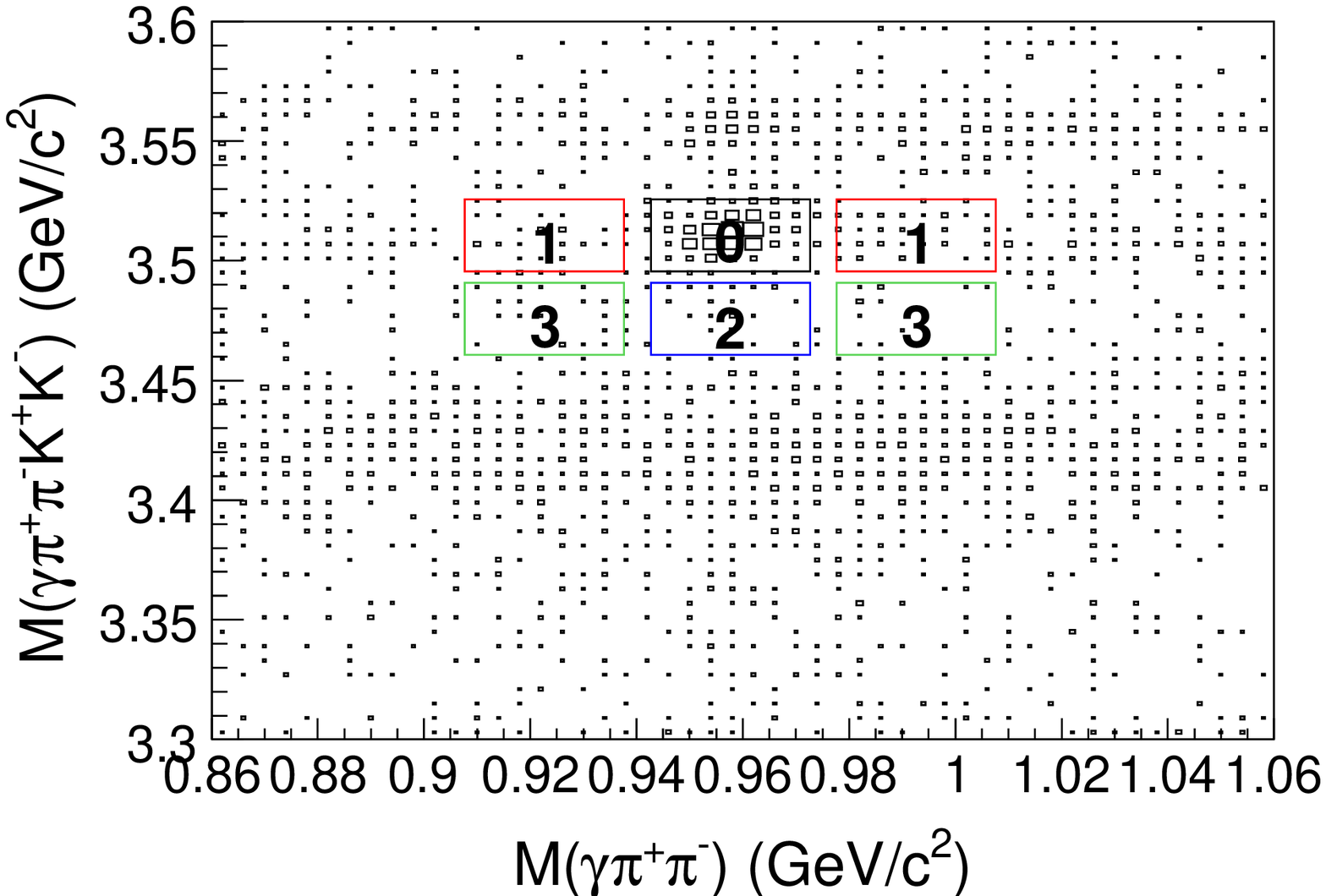}
\put(85,50){\large\bf (a)}
\end{overpic}
\begin{overpic}[width=8.0cm,height=5.0cm,angle=0]{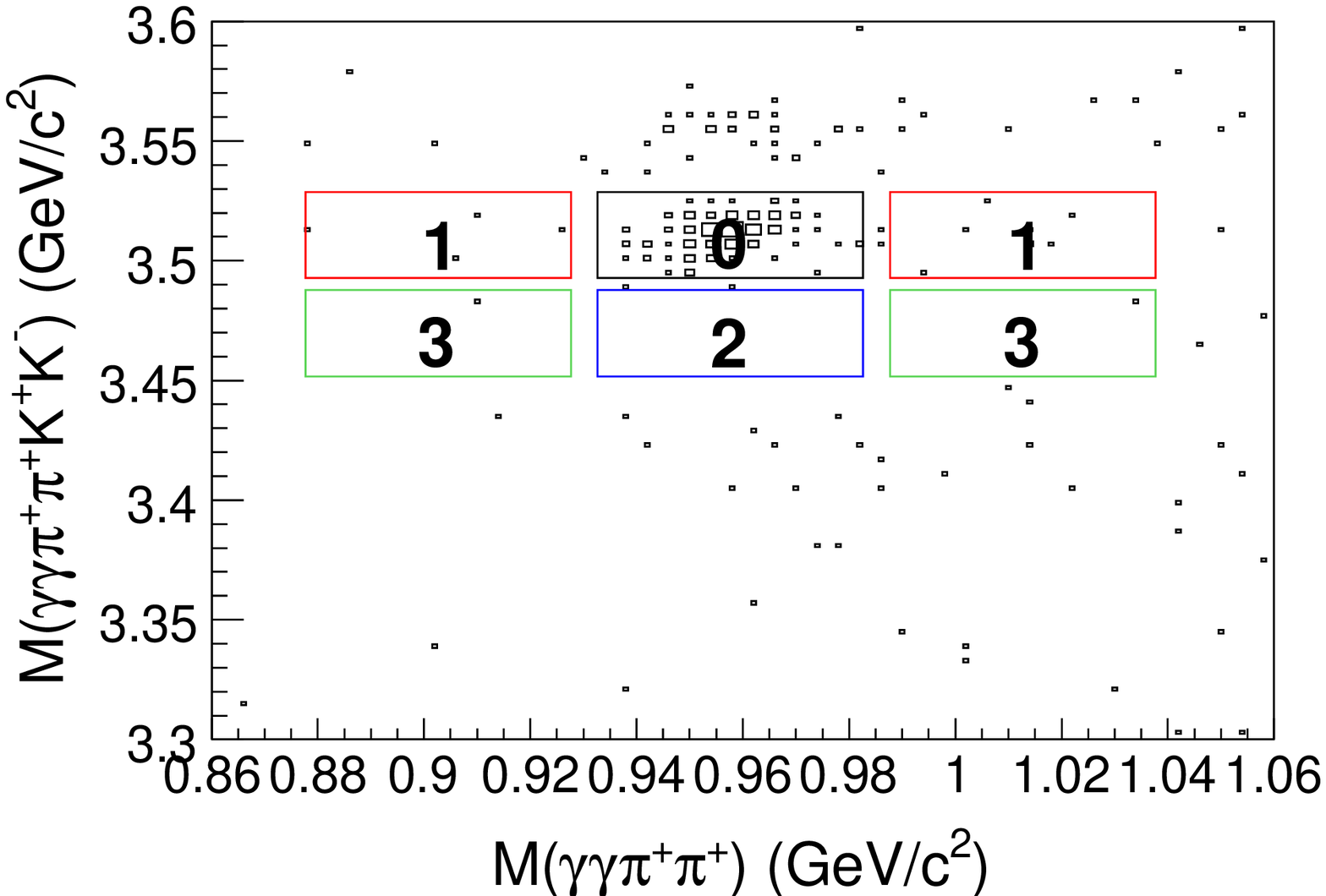}
\put(85,50){\large\bf (b)}
\end{overpic}
\parbox[1cm]{16cm} {\caption{(color online) (a) The scatter plot of
$M(\gamma\pi^+\pi^+K^+K^-)$ versus $M(\gamma\pi^+\pi^+)$ for mode
I. (b) The scatter plot of $M(\gamma\gamma\pi^+\pi^+K^+K^-)$ versus
$M(\gamma\gamma\pi^+\pi^+)$ for mode II. The plots here are the zoom-in
subregions of Fig.~\ref{2Detapsideband}(a) and Fig.~\ref{sidebandshape}(a)
around $\etap$ and $\chi_{cJ}$. The boxes defining the signal
and sideband regions are described in the text.}
\label{2Dsideband}
}
\end{figure*}

In the PWA, $\chicone$ signal candidate events are selected with
requirements I and III (box 0 in Fig.~\ref{2Dsideband}).  The first
category of background is the peaking $\gamma(\gamma)\pp\kk$
background in the $\chicone$ region, which is mainly from decay
processes with the same final states, or with one more (less) photon
in the final state, but without an $\etap$, the non-$\etap$
background.  This category of background can be estimated with events
within the $\etap$ sideband region with requirements II and III (boxes 1
in Fig.~\ref{2Dsideband}).  The second category of background is the
non-peaking background, the non-$\chicone$ background, which is mainly
from direct $\psi(3686)$ radiative decay,
$\psi(3686)\to\gamma\etap\kk$.  This background can be estimated with
the events within the $\chicone$ sideband region with requirements I
and IV (box 2 in Fig.~\ref{2Dsideband}).  There are also backgrounds
from processes without $\chicone$ and $\etap$ intermediate states, the
non-$\etap$ non-$\chicone$ background, which can be estimated with
events with requirements II and IV (boxes 3 in Fig.~\ref{2Dsideband}).
In the fit, background contributions to the log likelihood are
estimated from the weighted events in the sideband regions, and
subtracted in the fit, as following:
\begin{equation}
\begin{split}
 S =&\mathcal{S}_{sig}-\omega_{bkg1}\times\mathcal{S}_{bkg1}-\omega_{bkg2}\times\mathcal{S}_{bkg2}+\omega_{bkg3}\times \mathcal{S}_{bkg3}\\
   =& - \sum^{N_{sig}}_{i=1}\ln (\frac{\omega(\xi_i^k,\alpha)}
       {\int d\xi_i\omega(\xi_i^k,\alpha)\epsilon(\xi_i) })\\
    & + \omega_{bkg1} \times \sum^{N_{bkg1}}_{i=1} \ln (\frac{\omega(\xi_i^k,\alpha)}
       {\int d\xi_i\omega(\xi_i^k,\alpha)\epsilon(\xi_i)})\\
    & + \omega_{bkg2}\times \sum^{N_{bkg2}}_{i=1} \ln (\frac{\omega(\xi_i^k,\alpha)}
       {\int d\xi_i\omega(\xi_i^k,\alpha)\epsilon(\xi_i)})\\
    & - \omega_{bkg3}\times \sum^{N_{bkg3}}_{i=1} \ln (\frac{\omega(\xi_i^k,\alpha)}
       {\int d\xi_i\omega(\xi_i^k,\alpha)\epsilon(\xi_i)}),
\label{finallikelihood2}
\end{split}
\end{equation}
where $N_{sig}$, $N_{bkg1}$, $N_{bkg2}$ and $N_{bkg3}$ are the numbers
of events in the signal regions, non-$\etap$, non-$\chicone$ and
non-$\etap$ non-$\chicone$ sideband regions, respectively. The
$\omega_{bkg1}$, $\omega_{bkg2}$, and $\omega_{bkg3}$ are the
normalization weights of events in different sideband regions, and
are taken to be 0.5, 1.0, 0.5 in the fit, respectively. The sign before $\omega_{bkg3}$
is different with $\omega_{bkg1}$ and $\omega_{bkg2}$
because the third category of background is double counted in
the first two categories of background.

\subsection{\boldmath{PWA procedure and result}}
To improve the sensitivity for each sub-process, a combined fit on the
candidate events of the two $\etap$ decay modes is carried out, and the
combined log likelihood value:
\begin{equation}
\mathcal{S}_{total} = \mathcal{S}_1+\mathcal{S}_2= -\ln\mathcal{L}_{1}-\ln\mathcal{L}_{2}
\end{equation}
is used to optimize the fit parameters. Here, $\mathcal{S}_1$ and
$\mathcal{S}_2$ are the log likelihoods of the two decay modes,
respectively. In the fitting, two individual PHSP MC samples
($\psi(3686)\to\gamma\chi_{c1}$, $\chi_{c1}\to\etap\kk$,
$\etap\to\gamma\rho^0$ or $\etap\to\eta\pp$) are generated for the
normalized integral of the two $\etap$ decay modes,
respectively. Since the $\chi_{cJ}$ signal is included in the MC
samples, the propagator of $BW_{\chi_{cJ}}$ in Eq.~\ref{amp3} is set
to be unity in the fit.

Different combinations of states of $f_{0,2}(x)$,
$K^*_{0,1,2}(x)$ have been tested.  Because of the limited statistics,
 only the well established states in the PDG with statistical
significance larger than 5$\sigma$ are included in the nominal
result. Some different assumptions of the intermediate states are considered and will be described in detail in
section~\ref{checksolution}. Finally, only four intermediate states, $f_0(980)$, $f_0(1710)$, $f_2'(1525)$
and $K^*_0(1430)$, are included in the nominal result.

\begin{figure*}[htbp]
\centering
\begin{overpic}[width=16.0cm,height=4.0cm,angle=0]{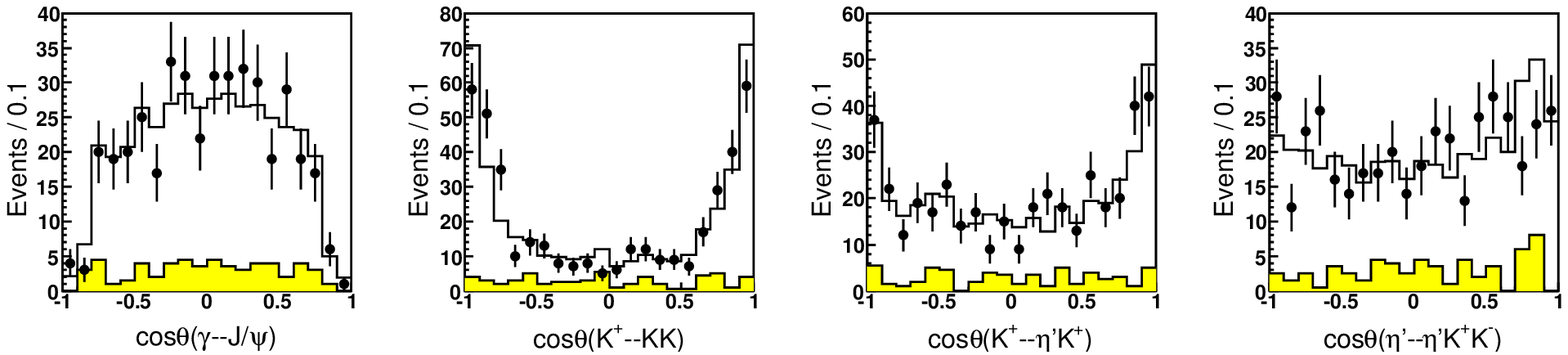}
\put(19,21){\large\bf (a)}
\put(44,21){\large\bf (b)}
\put(69,21){\large\bf (c)}
\put(94,21){\large\bf (d)}
\end{overpic}
\begin{overpic}[width=16.0cm,height=4.0cm,angle=0]{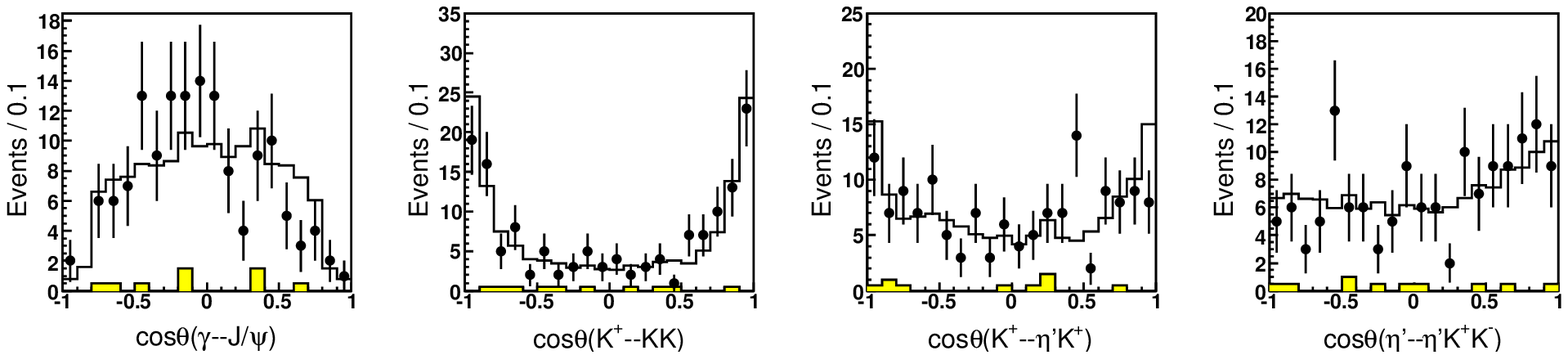}
\put(19,21){\large\bf (e)}
\put(44,21){\large\bf (f)}
\put(69,21){\large\bf (g)}
\put(94,21){\large\bf (h)}
\end{overpic}
\parbox[1cm]{16cm} {
\caption{(color online) Comparisons of angular distributions cos$\theta(\gamma-\jpsi)$, cos$\theta(K^{+}-K^{+}K^{-})$, cos$\theta(K^{+}-\etap K^{+})$,
 cos$\theta(\etap-\etap K^{+}K^{-})$, (a, b, c, d) for the $\etap$ decay mode I, (e, f, g, h) for the $\etap$ decay mode II. The empty histogram shows the global fit result combined with the background contribution. The filled histogram shows background.}
\label{project2}
}
\end{figure*}

\begin{figure}[htbp]
\centering
\begin{overpic}[width=8.0cm,height=4.0cm,angle=0]{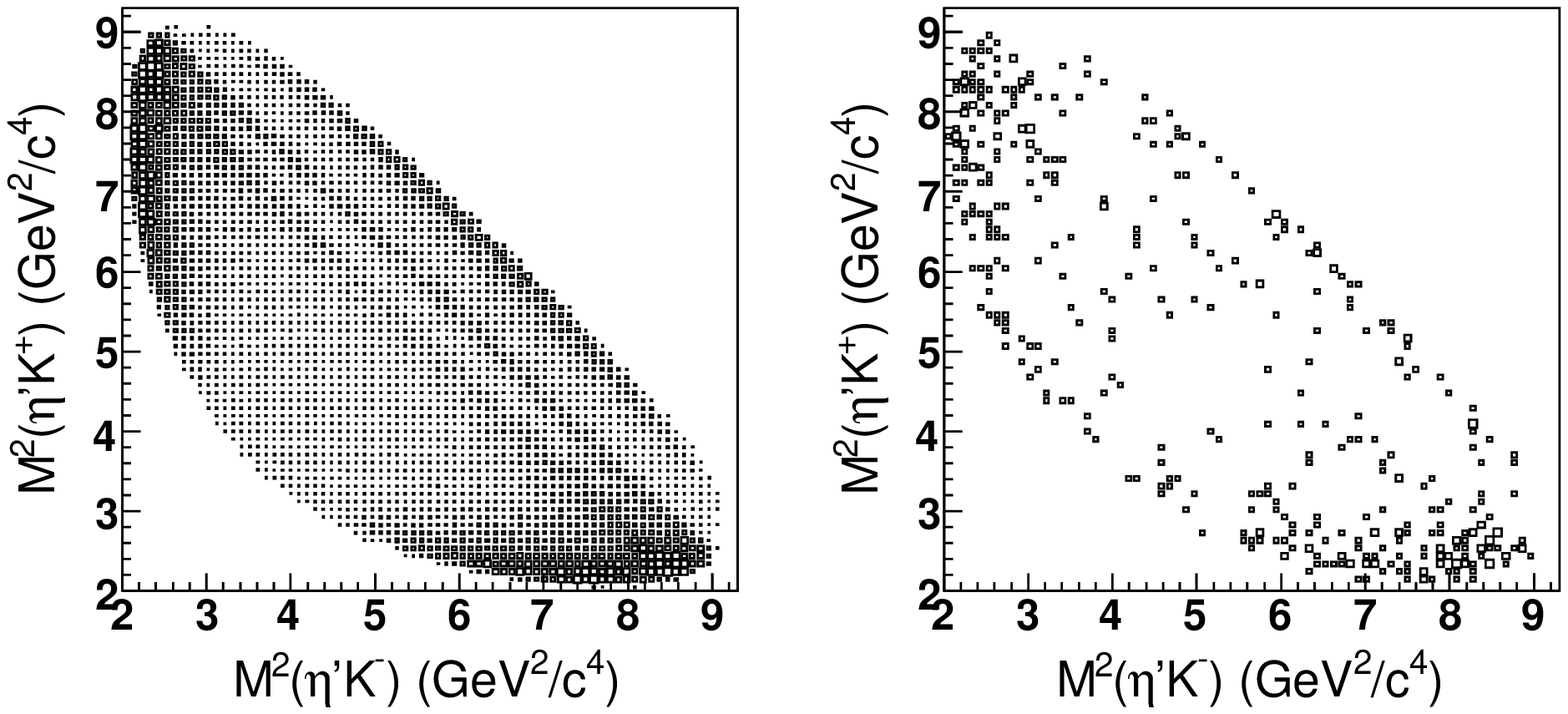}
\put(38,36){\large\bf (a)}
\put(88,36){\large\bf (b)}
\end{overpic}
\begin{overpic}[width=8.0cm,height=4.0cm,angle=0]{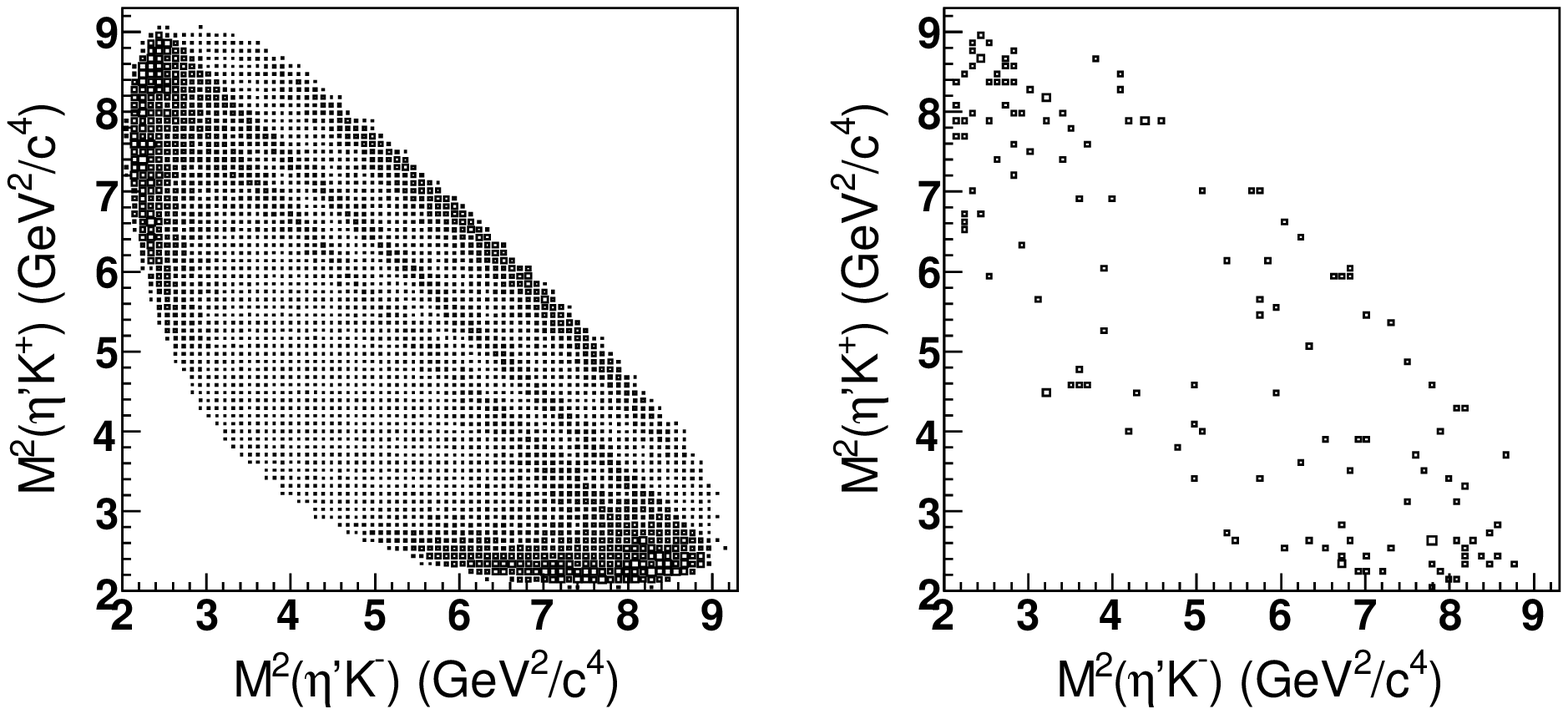}
\put(38,36){\large\bf (c)}
\put(88,36){\large\bf (d)}
\end{overpic}
\parbox[1cm]{8.6cm} {
\caption{Dalitz plots of $M^2(\etap K^+)$ versus $M^2(\etap K^-)$.
(a) of MC projections for the $\etap$ decay mode I;
(b) of data for the $\etap$ decay mode I;
(c) of MC projections for the $\etap$ decay mode II; and
(d) of data for the $\etap$ decay mode II.}
\label{dalitz}
}
\end{figure}

The $M(\kk)$ and $M(\gamma (\gamma)\pp K^\pm)$ distributions of data
and the PWA fit projections, as well as the contributions of individual
sub-processes for the optimal solution are shown in Fig.~\ref{project}
for the two $\etap$ decay modes.  The corresponding
comparisons of angular distributions $\theta(X-Y)$, the polar angle of
particle $X$ in $Y$-helicity frame, are shown in Fig.~\ref{project2}.
The PWA fit projection is the sum of the signal contribution of the best
solution and the backgrounds estimated with the events within the
sideband regions.  The Dalitz plots of data and MC projection from the
best solution of the PWA for the two $\etap$ decays modes are shown in
Fig.~\ref{dalitz}.

To determine goodness of the fit, a $\chi^2$ is calculated by
comparing data and the fit projection histograms, where $\chi^2$ is
defined as:
\begin{equation}
  \chi^2 =\sum^{r}_{i=1}\frac{(n_i-v_i)^2}{v_i} .
\end{equation}
Here $n_i$ and $v_i$ are the number of events for data and the fit
projections in the $i^{th}$ bin of each figure, respectively. If $v_i$
of one bin is less than five, the bin is merged to the
neighboring bin with the smaller bin content. The corresponding $\chi^2$
and the number of bins of each mass and angular distributions for the
two $\etap$ decay modes as well as for the combined distributions are
shown in Table~\ref{goodness}.  The values of $\chi^2/(N_{bin}-1)$ of
combined distributions are between 0.67 and 1.52, indicating
reasonable agreement between data and the fit projection.

\begin{table*}[htbp]
\centering
\caption{Goodness of fit check for the invariant mass and angular distributions.
\label{goodness}
}
\begin{tabular}{lcccccccc}\hline \hline
   & Variable     &~~~M$_{\kk}$&M$_{\etap K}$~~~
   &~~~$\theta_{\gamma-J\psi}$~~~ &~~~$\theta_{K^+-KK}$~~~ & ~~~$\theta_{K^{+}-\etap K^{+}}$~~~ & ~~~$\theta_{\etap-\etap K^{+}K^{-}}$~~~\\ \hline
   & $\chi^2$                     &56.6  &47.8 &10.8  &34.4  &20.1  &29.2  \\
$\etaptogammarho$~~~~~~~~& $N_{bin}$ &37    &46   &18    &20    &20    &20    \\
   & $\chi^2/(N_{bin}-1)$         &1.57  &1.06 &0.63  &1.81  &1.06  &1.54  \\ \hline
   & $\chi^2$                     &23.7  &74.3 &17.0  &6.6   &27.0  &20.4\\
$\etaptoetapipi$   & $N_{bin}$    &20    &33   &16    &14    &17    &20    \\
   & $\chi^2/(N_{bin}-1)$         &1.25  &2.32 &1.13  &0.51  &1.69  &1.07  \\ \hline
      & $\chi^2$                  &56.3  &59.9 &11.4  &27.2   &20.7  &17.7\\
Combine   & $N_{bin}$             &38    &46   &18    &20    &20    &20    \\
   & $\chi^2/(N_{bin}-1)$         &1.52  &1.33 &0.67  &1.43  &1.09  &0.93  \\ \hline\hline
\end{tabular}
\end{table*}

\subsection{\boldmath{Partial Branching fraction measurements}}
To get the branching fractions of individual sub-processes with
sequential two-body decay, the cross section fraction of the $i$th
sub-process is calculated with MC integral method:
\begin{equation}
  F_i = \sum_{j=1}^{N_{mc}} {(\frac{d\sigma}{d\phi})^i_j}/\sum_{j=1}^{N_{mc}}{(\frac{d\sigma}{d\phi})_j} .
\end{equation}
In practice, a large PHSP MC sample without any selection
requirements is used to calculate $F_i$, where
$(\frac{d\sigma}{d\phi})^i_j$ and $(\frac{d\sigma}{d\phi})_j$ are the
differential cross section of the $i$th sub-process and the total
differential cross section for the $j$th MC event, and $N_{mc}$ is
the total number of MC events.

The statistical uncertainties of the magnitudes,
phases and $F_i$ are estimated with a bootstrap
method~\cite{bootstrap}.  300 new samples are formed by random
sampling from the original data set; each with equal size as the
original.  All the samples are subjected to the same analysis as the
original sample. The statistical uncertainties of the magnitudes,
phases and $F_i$ are the standard deviations of the corresponding
distributions obtained and are listed in
Table.~\ref{basicsolution2}.

The partial branching fraction of the $i$th sub-process is:
\begin{equation}
\mathcal{B}_i = \mathcal{B}(\chicj\to\etap\kk) \times F_i
\end{equation}
where $\mathcal{B}(\chicj\to\etap\kk)$ is the average branching
fraction in Table~\ref{lastbrresult}.  The corresponding statistical
uncertainty of $\mathcal{B}_i$ contains two parts: one is from the
statistical uncertainty of $\mathcal{B}(\chicj\to\etap\kk)$
($\sigma_1$), and the other part is from the statistical uncertainty of
$F_i$ ($\sigma_2$).
\begin {eqnarray}
\begin{split}
\sigma_1 = \sigma(\mathcal{B}(\chicj\to\etap\kk)) \times F_i,&\\
\sigma_2 = \mathcal{B}(\chicj\to\etap\kk) \times \sigma(F_i),&\\
\end{split}
\label{extractstatistic}
\end {eqnarray}
The statistical uncertainty of
$\mathcal{B}(\chicj\to\etap\kk)$ is calculated with a weighted $\chi^2$
method:
\begin {eqnarray}
\begin{split}
\sigma(\mathcal{B}(\chicj\to\etap\kk))&=\sqrt{\frac{\sigma_{s1}^2\sigma_{s2}^2}{\sigma_{s1}^2+\sigma_{s2}^2}},&\\
\end{split}
\label{extractstatistic}
\end {eqnarray}
where $\sigma_{s1}$ and $\sigma_{s2}$ are the statistical
uncertainties given by the two decay modes listed in
Table~\ref{lastbrresult}. Finally the total statistical uncertainty of
the $i$th sub-process is:
\begin{equation}
 \sigma(\mathcal{B}_i)=\sqrt{\sigma_1^2+\sigma_2^2} .
\end{equation}
The results of cross section fraction $F_i$ and the partial branching
fractions of individual sub-processes as well as the two independent magnitudes and phase of each state of the baseline fit are shown in
Table~\ref{basicsolution2}, where only statistical uncertainties are
listed.
\begin {table*}[htp]
\begin {center}
\caption {The fitted magnitudes, phases, fractions and the
  corresponding partial branching fractions of individual processes
  in the nominal fit (statistical uncertainties only).
\label{basicsolution2}
}
\begin {tabular}{c|c|c|c|c|c} \hline \hline
\multirow{2}*{Process}&Magnitude  &Magnitude  &Phase                      &Fraction  &Partial Branching Fraction\\
\multirow{2}*{}       &$\rho_{i1}$&$\rho_{i2}$&$\phi_{i1}=\phi_{i2}$~(rad)&$F_i$~(\%)&$\mathcal{B} (10^{-4})$\\ \hline
\footnotesize{$\chicone\to K^*_0(1430)^{\pm}K^{\mp}, K^*_0(1430)^{\pm}\to\eta'K^{\pm}$}&1 (Fixed)    &$0.13\pm0.11$ &0 (Fixed) &$73.26\pm5.03$  &$6.41\pm0.57$ \\\hline
\footnotesize{$\chicone\to \eta'f_0(980), f_0(980)\to K^+K^-$}                        &$0.77\pm0.11$&$0.12\pm0.16$  &$5.50\pm0.28$ & $18.90\pm5.26$ &$1.65\pm0.47$ \\\hline
\footnotesize{$\chicone\to \eta'f_0(1710), f_0(1710)\to K^+K^-$}                      &$0.88\pm0.20$&$0.03\pm0.30$  &$0.96\pm0.18$ & $8.11\pm2.43$  &$0.71\pm0.22$ \\\hline
\footnotesize{$\chicone\to \eta'f_2'(1525), f_2'(1525)\to K^+K^-$}                    &$-0.17\pm0.03$&$0.01\pm0.05$  &$6.02\pm0.21$ & $10.50\pm2.63$ &$0.92\pm0.23$ \\
\hline \hline
\end {tabular}
\end {center}
\end {table*}

\subsection{\boldmath{Checks for the best solution}}{\label{checksolution}}
Various alternative PWA fits with different assumptions are carried
out to check the reliability of the results.  To get the statistical
significance of individual sub-processes, alternative fits with dropping one
given sub-process are performed. The changes of log
likelihood value $\Delta \mathcal{S}$ and of the number of degrees of
freedom $\Delta ndof$ as well as the corresponding statistical
significance are listed in Table~\ref{significancecheck}. Each sub-process has a statistical significance larger than 5$\sigma$.

\begin {table*}[htp]
\begin {center}
\caption { Change  in the log likelihood value $\Delta \mathcal{S}$,
  associated change of degrees of freedom $\Delta ndof$, and
  statistical significance if a process is dropped from the fit.
\label{significancecheck}
}
\begin {tabular}{cccccc} \hline \hline
Process & ~~$\chicone\to K^*_0(1430) K$~~   & ~~$\chicone\to f_{0}(980)\etap$~~&
          ~~$\chicone\to f_{0}(1710)\etap$~~& ~~$\chicone\to f_{2}'(1525)\etap$~~ \\ \hline
$\Delta \mathcal{S}$ & 323   & 89.7 &  22.8 &  33.2 \\
$\Delta {ndof}$      & 3     & 3    &  3    &  3    \\
Significance         & $\gg8\sigma$  & $\gg8\sigma$ & $6.2\sigma$ & $7.6\sigma$ \\ \hline \hline
\end {tabular}
\end {center}
\end {table*}

To determine the spin-parity of each intermediate state, alternative
fits with different spin-parity hypotheses of the $K^{\ast\pm}_X(1430)$,
$f_{X}(1710)$ and $f_{X}(1525)$ are performed.  If $J^{P}$ of
$K^{\ast\pm}_X(1430)$ is replaced with $1^-$ or $2^+$, the log
likelihood value is increased by 35 or 99, respectively. If $J^{PC}$
of $f_{X}(1525)$ is replaced with $0^{++}$, the log likelihood value
is increased by 12, while it increases by 7.4 when using the mass and
width of the $f_0(1500)$ in the fit. If $J^{PC}$ of $f_{X}(1710)$ is replaced with
$2^{++}$, the log likelihood value is improved by 1.3, so there is
some ambiguity for the $J^{PC}$ of the $f_{X}(1710)$ due to small statistics.
Since there is no known meson with $J^{PC}=2^{++}$ around $1.7
\gevcc$ in PDG, the structure around $1.7 \gevcc$ in $\kk$
invariant mass is assigned to be $f_0(1710)$ in the analysis. In the above
tests, the mass and width of each intermediate states are fixed to PDG values in the
fit~\cite{pdg}.  If we scan the mass and width of all the states,
$M(f_{X}(1710))\backsimeq1.705 \gevcc$ and
$\Gamma(f_X(1710))\backsimeq0.1331 \gevcc$, which agree well with
the PDG values, and the spin-parity of $f_{X}(1710)$ favors
$0^{++}$ over $2^{++}$ with log likelihood value improved by 11.

To check the contributions from other possible sub-processes,
alternative fits with additional known mesons listed in the PDG are
carried out.  Under spin-parity constraints, the intermediate mesons
$f_2(1270)$, $f_0(1370)$, $f_0(1500)$,
$f_2(1910)$, $f_2(1950)$, $f_2(2010)$, $f_0(2020)$, $f_0(2100)$, and
$f_2(2150)$ decaying to $\kk$, as well as  $K^*_1(1410)$,
$K^*_2(1430)$ and $K^*_1(1680)$ decaying to $\etap K^{\pm}$ are
included in the fit individually, and the masses and widths of these
intermediate states are fixed to values in the PDG. For $f_0(1370)$, there is
no average value in PDG, so its mass and width are fixed to the
middle value of the PDG range, $M=1.35\gevcc$, $\Gamma=0.35 \gevcc$.
To investigate
the contribution from the direct $\chicone\to\etap\kk$ decay (PHSP),
 two fits with different PHSP approximations are carried
out, where the first assumes that the $\kk$ system is a very broad
state with $J^{PC}=0^{++}$, and the other assumes that the $\etap
K^{\pm}$ system is a very broad state with $J^{P}=0^{+}$.  The
likelihood value change $\Delta \mathcal{S}$, the number of freedom
change $\Delta ndof$ as well as the corresponding significance of
various additional sub-process are summarized in
Table~\ref{additionaltest} and Table~\ref{additionaltest2}.
The sub-processes with intermediate state
of $f_0(2100)$, $K^*_2(1430)$ and $K^*_1(1680)$ have significances
larger than 5$\sigma$. $f_0(2020)$ has a significance of 4.9$\sigma$.
There might be some $f_0$ states around $2.1\gevcc$, but they are
not as well established as $f_0(1710)$ and $f_2'(1525)$, and it is
impossible to tell which might be here. Because they are far from
$f_0(1710)$ and should have little interference with other resonances,
we did not include any $f_0$ state around $2.1 \gevcc$ in nominal result.  Their
possible influence will be considered in the systematic uncertainty.
For $K^*_2(1430)$ and $K^*_1(1680)$, the large significance mainly
comes from the imperfect fit to real data with the $K^*_0(1430)$
lineshape cited.
If we scan the mass and width of intermediate states
in the fit instead of fixing them, the fit result agrees better with data
and the significances of the $K^*_2(1430)$ and $K^*_1(1680)$ are
only $0.6\sigma$ and $3.4\sigma$, respectively. It is therefore difficult
to confirm the existence of $K^*_2(1430)$ and
  $K^*_1(1680)$ decays to $K\etap$
with the available data, and these sub-processes are not included in
the nominal solution.  The influence on the measurement of these states
is considered in the systematic uncertainty. The fit results obtained
using resonance parameters from the mass and width scans are also
taken into account in the systematic uncertainty.

\begin {table*}[htp]
\begin {center}
\caption {The change of log likelihood value $\Delta \mathcal{S}$, of the number of freedom $\Delta ndof$
and the corresponding significance with additional processes on $K^+K^-$ invariant mass spectrum, where PHSP$_{1}$
represent for PHSP with $\kk$ broad states.
\label{additionaltest}
}
\begin {tabular}{ccccccccccc} \hline \hline
Add. res.            &$f_2(1270)$  &$f_0(1370)$ &$f_0(1500)$ & $f_2(1910)$ & $f_2(1950)$ & $f_2(2010)$   & $f_0(2020)$        &
                       $f_0(2100)$ & $f_2(2150)$ & PHSP$_{1}$   \\ \hline
$\Delta \mathcal{S}$ &6.0 & 10.2 &6.7 & 5.0         & 5.9         & 5.1           & 15.4        &
                       18.0        & 7.3         & 15.0         \\
$\Delta ndof$        &3   &3   & 3 &3          & 3           & 3             & 3           &
                       3           & 3           & 3            \\
Significance         &2.7$\sigma$ &3.8$\sigma$&2.9$\sigma$& 2.4$\sigma$ & 2.6$\sigma$ & 2.4$\sigma$   & 4.9$\sigma$ &
                       5.4$\sigma$ & 3.1$\sigma$ & 4.8$\sigma$  \\ \hline \hline
\end {tabular}
\end {center}
\end {table*}

\begin {table*}[htp]
\begin {center}
\caption {The change of log likelihood value $\Delta \mathcal{S}$, of the number of freedom $\Delta ndof$
and the corresponding significance with additional processes on $\eta'K$ invariant mass spectrum, where PHSP$_{2}$
represent for PHSP with $\eta'K$ broad states.
\label{additionaltest2}
}
\begin {tabular}{ccccc} \hline \hline
Add. res.             & $K^{\ast}_1(1410)$ & $K^{\ast}_2(1430)$ &
                       $K^{\ast}_1(1680)$  & PHSP$_{2}$ \\ \hline
$\Delta \mathcal{S}$ & 11.1        & 27.6        &      19       & 15.0          \\
$\Delta ndof$        & 3           & 3           & 3             & 3             \\
Significance         & 4.0$\sigma$ & 6.8$\sigma$ &   5.7$\sigma$ & 4.8$\sigma$   \\ \hline \hline
\end {tabular}
\end {center}
\end {table*}

\subsection{\boldmath{The systematic uncertainty}}
Several sources of systematic uncertainty are considered in
determination of the individual partial branching fractions:

\emph{a. The value of the centrifugal barrier R~~} In the fit, centrifugal
barrier R is 1.0 fm.  Alternative PWA fits with R varied from 0.1 fm
to 1.5 fm are performed.  The differences of partial branching fractions from
the nominal results are taken as the systematic uncertainties from the
centrifugal barrier.

\emph{b. The uncertainty from additional states~~} As mentioned above,
there are possible contributions from other sub-processes with
different intermediate states in $\chicone\to\etap \kk$ decay.
Several alternative fits including known states listed in the PDG and
the two different approximation of PHSP are carried out, and the
largest differences of partial branching fractions are taken as the systematic
uncertainties.

\emph{c. The shape of $K^\ast_0(1430)$~~} Because
$K^\ast_0(1430)$ is at the $\etap K^\pm$ threshold, the
Flatt\'{e} formula (Eq. \ref{cleoflatte}) is used to parameterize the
shape of $K^\ast_0(1430)$ in nominal fit.  A PWA with an alternative
Flatt\'{e} formula:
\begin {equation}
\begin{split}
\label{flatte05}
f(s)=&\frac{1}{M^2-s-iM\Gamma(s)},\\
\Gamma(s)=&\frac{s-s_{A}}{M^2-s_{A}}\cdot g_1^2\cdot\rho_{K\pi}(s)+\frac{s-s_{A}}{M^2-s_{A}}\cdot g_2^2\cdot \rho_{K\eta'}(s),\\
\end{split}
\end {equation}
for $K^\ast_0(1430)$ is performed.  Here $M=1.517\gevcc$, the
Adler zero $S_A=m_{K}^2-m_\pi^2/2\simeq0.23\,\mathrm{GeV}^2/c^4$,
$g_1^2=0.353\gevcc$, and $g_2^2/g_1^2=1.15$, are from
Ref.~\cite{buggkstar}.  As mentioned at the end of
section~\ref{checksolution}, the fit result using resonance parameters from the mass and width scans
 are also considered. The largest differences of the partial branching fractions  to
the nominal values are taken as the systematic uncertainties
associated with the $K^\ast_0(1430)$ parameterization.

\emph{d. The mass and width uncertainties of intermediate states~~} As mentioned in section ~\ref{bwform}
, the mass and width of intermediate states, i.e. $f_0(1710)$,
$f_2'(1525)$ and $K^\ast_0(1430)$ are fixed to the values in the PDG
or in the corresponding literature.  PWA fits with changes in the masses and widthes
of intermediate states by 1$\sigma$ are performed individually. The
largest differences on the partial branching fractions are taken as the
systematic uncertainties.

\emph{e. Background uncertainty~~} To estimate the systematic
uncertainty from background, alternative intervals of sideband regions
are defined, and the PWA fit is redone. The differences to the
nominal partial branching fractions are taken as the systematic uncertainties.

\emph {f. The uncertainty from
$\mathcal{B}(\chi_{c1}\to\eta'K^+K^-)$~~} Because the total branching
fraction $\mathcal{B}(\chi_{cJ}\to\eta'K^+K^-)$ is used to calculate
the individual partial branching fractions of intermediate states, the systematic
uncertainty of $\mathcal{B}(\chi_{cJ}\to\eta'K^+K^-)$, $0.75\times
10^{-4}$, must be included.

A summary of the partial branching fraction systematic uncertainties for
individual sub-processes are shown in Table~\ref{sumpwabr}.  The total
systematic uncertainties are obtained by adding the individual
contributions in quadrature.
\begin {table*}[htp]
\begin {center}
\caption {Summary for systematic uncertainties of partial branching fraction
  of intermediate states (in \%).
\label{sumpwabr}
}
\begin {tabular}{lcccc} \hline \hline
                                  & ~~$K^*_0(1430)$~~   & ~~$f_0(980)$~~    & ~~$f_0(1710)$~~   & ~~$f_2'(1525)$~~ \\ \hline
The R Value                       & $^{+2.0}_{-9.1}$    & $^{+12.6}_{-12.0}$&$^{+18.0}_{-23.6}$ & $^{+12.9}_{-28.0}$  \\
The additional states             & $^{+22.2}_{-40.4}$  & $^{+58.7}_{-25.7}$&$^{+93.1}_{-54.2}$ & $^{+51.6}_{-39.8}$  \\
The shape of $K_0^{\ast}(1430)$   & $^{+22.2}_{-0}$     & $^{+52.1}_{-0}$   &$^{+0}_{-26.4}$    & $^{+26.1}_{-0}$  \\
The background                    & $^{+0}_{-0.2}$      & $^{+0}_{-16.7}$   &$^{+0}_{-15.5}$    & $^{+0}_{-23.9}$  \\
Mass\&width uncertainty on PDG    & $^{+1.4}_{-0.9}$    & $^{+4.8}_{-1.8}$  &$^{+4.2}_{-4.2}$   & $^{+2.2}_{-1.1}$ \\
$\mathcal{B}(\chi_{cJ}\to\eta'K^+K^-)$     & $^{+8.6}_{-8.6}$    & $^{+8.6}_{-8.6}$  &$^{+8.6}_{-8.6}$   & $^{+8.6}_{-8.6}$ \\\hline
Total                             & $^{+32.6}_{-42.3}$  & $^{+80.1}_{-34.1}$& $^{+95.3}_{-67.3}$& $^{+59.9}_{-54.9}$       \\ \hline \hline
\end {tabular}
\end {center}
\end {table*}

\section{\boldmath PWA for $\chictwo$}
\begin{figure*}[htbp]
\centering
\begin{overpic}[width=0.9\linewidth,height=0.35\linewidth,angle=0]{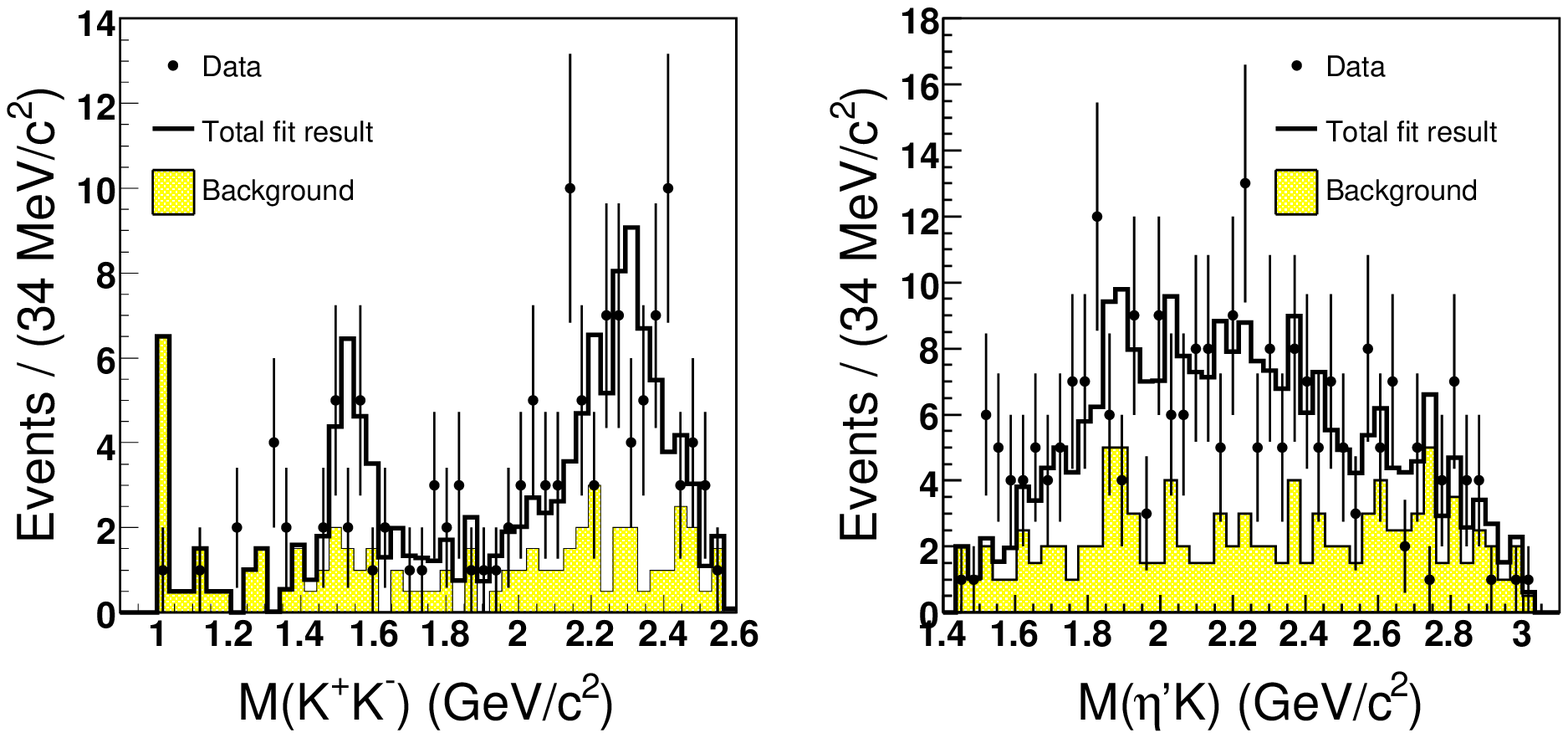}
\put(26,25){\large\bf (a)}
\put(92,25){\large\bf (b)}
\end{overpic}
\begin{overpic}[width=0.9\linewidth,height=0.35\linewidth,angle=0]{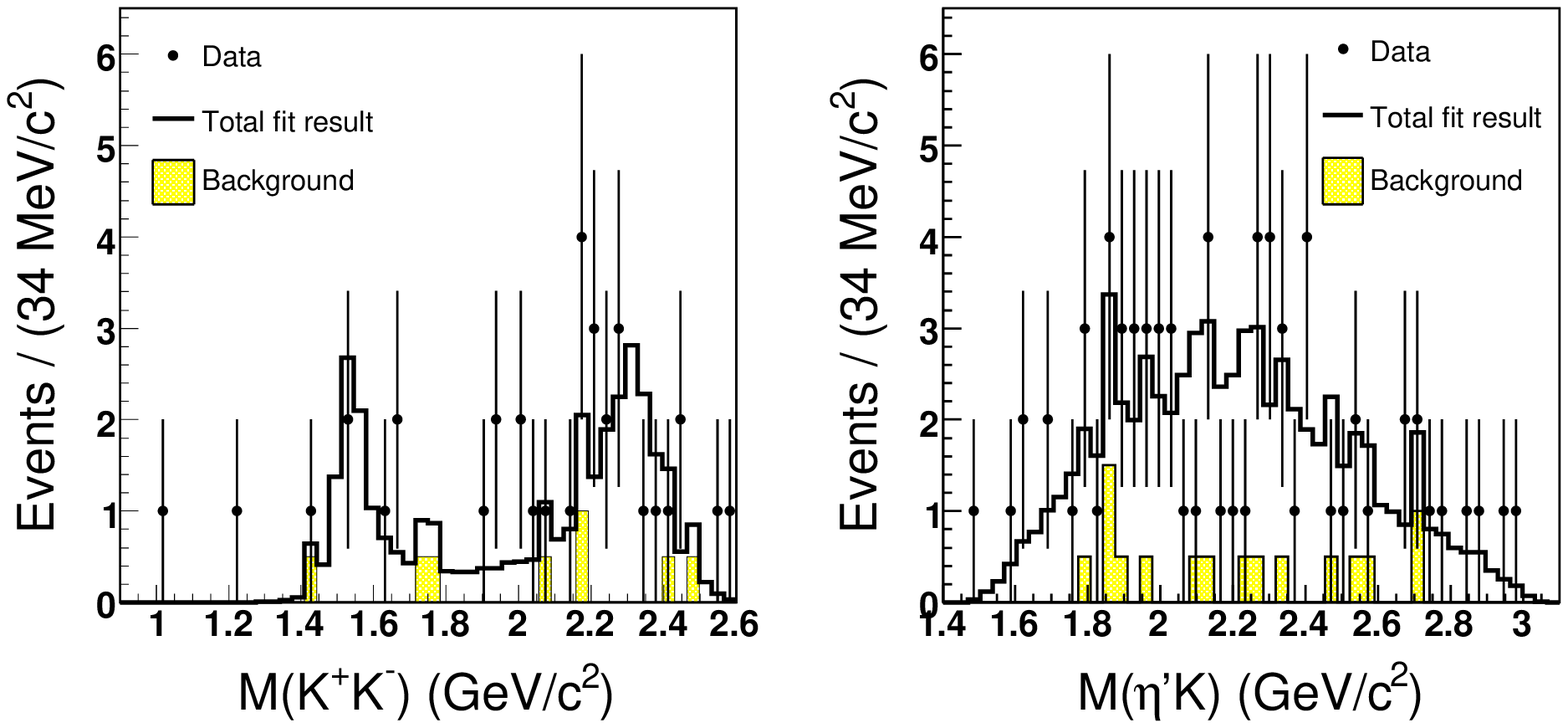}
\put(26,25){\large\bf (c)}
\put(92,25){\large\bf (d)}
\end{overpic}
\parbox[1cm]{16cm} {
\caption{(color online) The invariant mass distributions of $\kk$ and $\gamma
  (\gamma)\pp K^\pm$ for events within the $\chictwo$ selection range.
(a)(b) for the $\etap$ decay mode I, and (c)(d) for the $\etap$ decay mode II.}
\label{projectchic2}
}
\end{figure*}

Fig.~\ref{projectchic2} shows the $M(\kk)$ and $M(\gamma (\gamma)\pp
K^\pm)$ distributions after the $\chictwo$ mass window requirement:
$|M(\gamma\pp\kk)-M(\chictwo)|<16\mevcc$ for mode I and
$|M(\gamma\gamma\pp\kk)-M(\chictwo)|< 18 \mevcc$ for mode II.  There
is a small structure around $1.5 \gevcc$ and a very wide structure
around $2.3 \gevcc$ in the $\kk$ invariant mass spectrum.  No obvious
structure is observed in the $\etap K^\pm$ invariant mass spectrum.
From spin-parity conservation, the decays $\chictwo\to f_0 \etap$ and
$\chictwo\to K^{\ast \pm}_{0} K^\mp$ are forbidden. A possible process
is $\chictwo\to f_2 \etap$.  Since there are few events and the
background is about 50\%, estimated by fitting of $\etap \kk$ invariant mass
distribution, a simple simultaneous PWA fit is performed on the
candidate events of the two $\etap$ decay modes.  No intermediate
state results are given; the PWA is only used to generate MC samples
to determine the detection efficiency of $\chictwo\to\etap\kk$.

In the PWA, only $f_2'(1525)$ and $f_2(2300)$ states in the $\kk$
invariant mass distribution are considered. The mass and width of
$f_2'(1525)$ are fixed to PDG values~\cite{pdg}. The mass and width of
$f_2(2300)$ are about $2.323 \gevcc$ and $0.183 \gevcc$ from a rough scan.  The PWA fit
with or without background subtraction is performed, where the background
is estimated from the $\etap$ sideband events. The difference of
detection efficiency given for the two cases is taken as systematic
uncertainty when measuring $\mathcal{B}(\chi_{c2}\to\etap\kk)$.

\section{Summary}
Based on a sample of $(106.41\pm 0.86) \times 10^{6}$ $\psi(3686)$ events
collected with the BESIII detector, the branching fractions of
$\chi_{c1,2}\to\etap\kk$ are measured with $\etap\to\gamma\rho^{0}$
and $\etap\to\eta\pp$. The measured branching fractions are summarized
in Table~\ref{lastbrresult}.  Abundant structures on the
$\kk$ and $\etap K^{\pm}$ invariant mass spectra are observed for
$\chicone$ candidate events, and a simultaneous PWA
with covariant tensor amplitudes is performed for the two $\etap$ decay modes.  The partial branching fractions of
$\chicone$ decay processes with intermediate states $f_0(980)$,
$f_0(1710)$, $f_2'(1525)$ and $K^{\ast}_0 (1430)$ are measured and
summarized in the Table~\ref{lastbrresult}.
\begin {table*}[htp]
\begin {center}
\caption {The branching fractions of $\chi_{c1,2}\to\etap\kk$ and partial branching fractions of
$\chi_{c1}$ decay to intermediate states.  The first uncertainties
are statistical, and the second are systematic.  For the average
branching fraction, the uncertainty is the combined uncertainty.
\label{lastbrresult}
}
\begin {tabular}{lcc}
\hline\hline
\multicolumn{2}{l}{Process } &  $\mathcal{B}(\times10^{-4})$    \\
\hline
\multirow{3}*{$\mathcal{B}(\chicone\to\etap\kk)$} &$\etap\to\gamma\rho^0$  &$9.09\pm0.54\pm0.86$ \\
\multirow{3}*{}                                   &$\etap\to\eta\pi^+\pi^-$  &$8.33\pm0.77\pm0.77$ \\
\multirow{3}*{}                                   &average   &$8.75\pm0.87$ \\\hline
\multirow{3}*{$\mathcal{B}(\chictwo\to\etap\kk)$} &$\etap\to\gamma\rho^0$   &$1.84\pm0.31\pm0.33$ \\
\multirow{3}*{}                                   &$\etap\to\eta\pi^+\pi^-$    & $2.05\pm0.41\pm0.25$\\
\multirow{3}*{}                                   &average   & $1.94\pm0.34$\\\hline
\multicolumn{2}{l}{\footnotesize{$\chi_{c1}\to K^*_0(1430)^{\pm}K^{\mp}, K^*_0(1430)^{\pm}\to\eta'K^{\pm}$}}~~~~&~~~~$6.41\pm0.57^{+2.09}_{-2.71}$~~~~ \\
\multicolumn{2}{l}{\footnotesize{$\chi_{c1}\to \eta'f_0(980), f_0(980)\to K^+K^-$}}& $1.65\pm0.47^{+1.32}_{-0.56}$ \\
\multicolumn{2}{l}{\footnotesize{$\chi_{c1}\to \eta'f_0(1710), f_0(1710)\to K^+K^-$}}& $0.71\pm0.22^{+0.68}_{-0.48}$ \\
\multicolumn{2}{l}{\footnotesize{$\chi_{c1}\to \eta'f_2'(1525), f_2'(1525)\to K^+K^-$}}& $0.92\pm0.23^{+0.55}_{-0.51}$ \\ \hline
\hline
\end {tabular}
\end {center}
\end{table*}
All of these branching fractions are measured for the first time.  As
mentioned in the introduction, the results can be used to constrain
glueball-$q\overline{q}$ mixing schemes for scalar mesons.
However, both the theory in reference~\cite{wangqian} and our measurement result
has large uncertainty. Our result can
not distinguish between the mixing schemes. The decay $K^*_0(1430)^{\pm}\to\etap
K^{\pm}$ is observed for the first time.

\section{Acknowledgments}
The BESIII collaboration thanks the staff of BEPCII and the computing center for their strong support. This work is supported in part by the Ministry of Science and Technology of China under Contract No. 2009CB825200; Joint Funds of the National Natural Science Foundation of China under Contracts Nos. 11079008, 11179007, U1332201; National Natural Science Foundation of China (NSFC) under Contracts Nos. 10625524, 10821063, 10825524, 10835001, 10935007, 10979038, 11005109, 11079030, 11125525, 11235011, 11275189, 11322544, 11375204; the Chinese Academy of Sciences (CAS) Large-Scale Scientific Facility Program; CAS under Contracts Nos. KJCX2-YW-N29, KJCX2-YW-N45; 100 Talents Program of CAS; German Research Foundation DFG under Contract No. Collaborative Research Center CRC-1044; Istituto Nazionale di Fisica Nucleare, Italy; Ministry of Development of Turkey under Contract No. DPT2006K-120470; U. S. Department of Energy under Contracts Nos. DE-FG02-04ER41291, DE-FG02-05ER41374, DE-FG02-94ER40823, DESC0010118; U.S. National Science Foundation; University of Groningen (RuG) and the Helmholtzzentrum fuer Schwerionenforschung GmbH (GSI), Darmstadt; WCU Program of National Research Foundation of Korea under Contract No. R32-2008-000-10155-0.

\end{document}